\documentclass[aps,amsmath,amssymb,reprint,a4paper]{revtex4-2}
\usepackage[utf8]{inputenc}
\usepackage[english]{babel}
\usepackage{graphicx}
\usepackage{amsfonts,amsthm}
\usepackage{color, colortbl}
\usepackage{hyperref}
\hypersetup{colorlinks=true,allcolors=blue}
\usepackage{hypcap}
\usepackage{subcaption} 
\usepackage{tkz-euclide}

\usepackage{rotating}
\usepackage{makecell}

\definecolor{Gray}{gray}{0.95}
\newtheorem{assumption}{Assumption}
\newtheorem{example}{Example}

\begin{document}

\preprint{APS/123-QED}


\title{A normal form for grid forming power grid actors.}
\author{Raphael Kogler}
\author{Anton Plietzsch}
\author{Paul Schultz}
\author{Frank Hellmann}
\email[]{hellmann@pik-potsdam.de}
\affiliation{Potsdam Institute for Climate Impact Research, Telegrafenberg A31, 14473 Potsdam, Germany}

\begin{abstract}

Future power grids will be operating a large number of heterogeneous dynamical actors. Many of these will contribute to the fundamental dynamical stability of the system, and play a central role in establishing the self-organized synchronous state that underlies energy transport through the grid. By taking a complexity theoretic perspective we derive a normal form for grid forming components in power grids. This allows analyzing the grids systemic properties in a technology neutral manner, without detailed component models.

Our approach is based on the physics of the power flow in the grid on the one hand, and on the common symmetry that is inherited from the control objectives grid-forming power grid components are trying to achieve. We provide a first experimental validation that this normal form can capture the behavior of complex grid forming inverters without any knowledge of the underlying technology, and show that it can be used to make technology independent statements on the stability of future grids.

\end{abstract}


\maketitle 

\section*{Introduction}

The transport of energy through the power grid depends on a self-organized synchronous state of distributed dynamical actors at continental scale. These dynamical actors are called grid-forming. They establish the dynamical state on which power flow is possible. In the current grid these are primarily conventional power plants, with control schemes designed around heavy rotating masses. The energy transition demands a shift from such conventional generators towards inverter-interfaced renewable energy sources, with their dynamics specified by power electronics. This poses a fundamental challenge to understand the collective phenomena of these novel grid-forming components, and ensure the existence and resilience of the self-organized synchronous state. Consequently, the collective dynamics of power grids has become a very active interdisciplinary area of research in recent years.


The design of conventional generators is well established and grounded in the physics of synchronous machines. There is a good understanding in the engineering community what level of model detail is required to study which questions. This is not the case for inverter-interfaced energy sources. Here the correct design of the control is an ongoing topic of research, e.g. \cite{chen2011improving, chen2012comparison, schiffer2014, doerfler2018}, especially for so called grid-forming inverters, which, unlike grid-following inverters, do not rely on the pre-existence of a stable grid. Consequently there is, as of yet, no clear consensus on the proper dynamical modeling of such future sources of energy.

This situation is especially problematic as we expect the number of dynamical actors in a future power grid to increase by orders of magnitude, as energy generation is going to move more and more from the transmission to the distribution level. Further, we expect a larger heterogeneity in dynamics, as the control is no longer structured around established principles, i.e. principles dictated by the physical components. Consequently, it can not be expected that all the inverters participating in future power grids will use similar control designs. At the same time, the future power grid will face numerous dynamic stability issues such as low inertia~\cite{doerfler2018a}, (lack of)  time scale separation~\cite{doerfler2019} and greater fluctuations of power production~\cite{haehne2018, haehne2019} and consumption~\cite{muratori2018, anvari2020}. The increasing number of dynamical actors and, in turn, heterogeneity mean that, in tackling these issues, we need to understand not only the intrinsic dynamics of individual actors, but more importantly their interactions in large complex networks.

This paper addresses these challenges by providing a complexity theoretic approach towards the modeling of future power grids. This means that rather than starting with detailed models of the individual actors and subsequently simplifying them, we focus on those features that are crucial for their interaction on the network. Our approach is based on the physical relationship between voltage and current, which provides the coupling on the network, on the one hand, and on the system desiderata and the symmetry implied by them on the other hand. We will find that the most important desiderata are active and reactive power injection, and prescribed voltage levels.

Formulating the dynamics of the nodes in the power grid in terms of the natural invariants associated with their symmetry, we can give an order by order nonlinear approximation of their behavior in terms of the deviations from the local desiderata. The key insight is that, by choosing appropriate variables, the lowest non-trivial order in this approximation is able to capture the most fundamental nonlinearities of the power grid dynamics, while also being capable of expressing all local desiderata. The result is a normal form for the behavior of grid-forming actors that resembles controlled Stuart-Landau oscillators, analogous to the form that appears in bifurcation theory~\cite{kuznetsov1998}, parametrized by a latent linear input-output system. From a dynamical systems perspective the crucial point here is that we fix not only the dynamics of the oscillators to be of Stuart-Landau type, but also provide a specific form for their coupling.

We show by means of comparison to experimental measurements from a sophisticated inverter implemented in the lab, as well as more broadly through numerical simulations, that our normal form indeed captures the quantitative and qualitative properties of a wide range of actors. We also find that it captures important features of the nonlinear interaction that occurs in highly heterogeneous power grids containing both conventional generators and grid-forming inverters.

This normal form provides a starting point for studying realistic models of future power grids from a truly transdisciplinary perspective. By providing a form that closely resembles the Stuart-Landau oscillator it enables the application of a large range of dynamical systems results in the context of power grids. The fact that the normal form is parametrized by a latent linear input-output system enables the use of tools like model reduction and system identification. As opposed to the phase models (e.g. the Swing equation) used in much of the control theoretic and complex systems literature on power grids, the normal form here is based on the instantaneous, physically relevant variables, and provides, order by order, all relevant dynamical aspects of the dynamical actors described.

\section*{Results}

\subsection{Modeling power grids}\label{sec:modeling}

We start by introducing our approach to the basic modeling of power grids as complex dynamical networks. This section serves a dual purpose. To make the paper self-contained to researchers without a strong background in power system modeling we briefly introduce the most salient physical properties of power grids. We then use this introduction to spell out a sequence of fairly general assumptions we make on our nodes in order to arrive at a highly generic model of power grid components. For a more extensive background on overall power-grid and inverter modeling we refer the reader to~\cite{schiffer2016, doerfler2017} (and the references therein), which we largely follow in this regard.

Our overall approach is to consider nodes as specifying a voltage which in turn causes a current to flow on the power lines. In electrical engineering terms we thus think of nodes as capacitive elements, and consider our dynamical components as controlled voltage sources. While many of the concrete assumptions below are highly general and apply to all types of power grid components, the assumption of capacitive behavior is most natural for grid forming voltage source inverters.

\paragraph*{Notational aside:} In accordance with mathematical and dynamical systems literature, we use $i = \sqrt{-1}$ to denote the imaginary unit and use $j$ for the current (this is the opposite convention to the electrical engineering literature).

\subsubsection{Node model}

Most AC power grids in operation today are three-phase, i.e. at every node in the network there are three alternating voltages, $V_a(t), V_b(t), V_c(t): \mathbb{R}\to\mathbb{R}$, which cause three separate alternating currents on each transmission element. The sum of the currents flowing from a node into the grid is then denoted $I_a(t), I_b(t), I_c(t): \mathbb{R}\to\mathbb{R}$. Our first assumption is that the three phases of both voltage and current are balanced (symmetrical) at all times. That is, we assume $V_a(t) + V_b(t) + V_c(t) = 0$ and $I_a(t) + I_b(t) + I_c(t) = 0$ for all $t$. This allows us to represent them each by a complex variable with the aid of what is called the Clarke or $\alpha\beta$ transformation\cite{duesterhoeft1951determination}. We denote the complex nodal voltage by $u(t):\mathbb{R}\to\mathbb{C}$, with $V_a = \sqrt{2/3} ~\Re(u)$, $V_{b,c} = \sqrt{2/3} ~\Re\left( u \exp(\pm 2\pi i/3 ) \right)$, and the complex nodal current by $j(t):\mathbb{R}\to\mathbb{C}$, with $I_a = \sqrt{2/3} ~\Re(j)$, $I_{b,c} = \sqrt{2/3} ~\Re\left( j \exp(\pm 2\pi i/3 ) \right)$.

\begin{assumption}[Balanced phases]
The nodal voltage and current are balanced at all times and can thus be described in terms of the complex variables $u$ and $j$ respectively.
\end{assumption}

The main objective of AC power grid control is to reach and maintain an operating state where all nodal voltages and currents rotate at a uniform frequency $\Omega_s$ (usually 50 or 60~Hz), i.e. a quasi-steady-state described by a limit cycle where at each node we have a nodal voltage $u_s(t) \sim \exp(i\Omega_s t)$ and current $j_s(t) \sim \exp(i\Omega_s t)$. This operating state is further determined by specifying certain set-points, i.e. desired values for the amplitude of the nodal voltage, $\rho_s := \lvert u_s \rvert$, and active as well as reactive power input, $p_s := \Re(u_s j_s^*)$ and $q_s := \Im(u_s j_s^*)$ respectively. These set-points must be provided by some higher-level control in accordance with the desired power flow in the grid. Since we are primarily interested in the dynamics of the fast acting primary control that takes place at the sub-second scale, we assume that the control dynamics can be described as fully decentralized, i.e. we assume that the set-points are given constants and, furthermore, we preclude any additional communication between the individual nodes such that information about the state of the power grid can only be inferred from local measurements. Since we consider controllable voltage sources, this means that the dynamical coupling between the nodes of the network can only be realized by measurements of the local nodal current.

\begin{assumption}[Decentralized control]
a) The set-points specifying the desired operating state are given constants and b) the dynamics at a node depend on the other nodes only via the currents on the transmission lines.
\end{assumption}

While the collective network dynamics with this form of coupling can thus be described in terms of the nodal voltages and currents, an individual node may also have any number of internal dynamic variables. We assume that these are either balanced three-phase quantities (internal three-phase voltages and currents) or scalar quantities (e.g. frequency, DC voltages and currents, or auxiliary variables). For a particular component featuring $L$ of the former and $K$ of the latter, we denote them by $z(t):\mathbb{R}\to\mathbb{C}^L$ and $x(t):\mathbb{R}\to\mathbb{R}^K$ with their respective operating states $z_s(t)\sim \exp(i\Omega_s t)$ and constant $x_s$. This distinction is similar to that between AC and DC variables in \cite{doerfler2017} with the difference that we do not explicitly model phase variables.

\begin{assumption}[Internal variables]
The node dynamics may feature any number of internal variables comprising either balanced three-phase variables denoted by $z$ or scalar variables denoted by $x$.
\end{assumption}

Next, we assume that the nodal voltage as well as the internal variables react smoothly on a given nodal current and can be formulated as a system of ordinary differential equations with input $j(t)$.

\begin{assumption}[Smooth dynamics]
The node dynamics is smooth and can be formulated as a system of ordinary differential equations in terms of $u$, $z$, and $x$, with an input given by $j$.
\end{assumption}

Putting all of these assumptions together, the general form of the node dynamics that we are considering here is given by the system
\begin{equation}\begin{aligned}\label{eq:1-oscillator}
	\dot{u} &= f^u(x, z, z^*, u, u^*, j, j^*) \;,\\
	\dot{z} &= f^z(x, z, z^*, u, u^*, j, j^*) \;,\\
	\dot{x} &= f^x(x, z, z^*, u, u^*, j, j^*) \;,
\end{aligned}\end{equation}
where $f^u: \mathbb{R}^{K} \times \mathbb{C}^{L + 2} \to \mathbb{C}$, $f^z: \mathbb{R}^{K} \times \mathbb{C}^{L + 2} \to \mathbb{C}^{L}$, and $f^x: \mathbb{R}^{K} \times \mathbb{C}^{L + 2} \to \mathbb{R}^{K}$ are some smooth functions in the sense that their real and imaginary parts are differentiable functions of the real and imaginary parts of $u$ and $j$. From this it follows that they can be written as holomorphic functions of $u$, $u^*$, $j$ and $j^*$, e.g. \cite{remmert1991theory}. The equations \eqref{eq:1-oscillator} admit solutions congruent with the required operating state, i.e. for some operating state given by $u_s$ and $j_s$ there exist $z_s$ and $x_s$ for which
\begin{equation}\begin{aligned}\label{eq:qss}
	f^u(x_s, z_s, z^*_s, u_s, u^*_s, j_s, j^*_s) &= i\Omega_s u_s \;,\\
	f^z(x_s, z_s, z^*_s, u_s, u^*_s, j_s, j^*_s) &= i\Omega_s z_s \;,\\
	f^x(x_s, z_s, z^*_s, u_s, u^*_s, j_s, j^*_s) &= \vec{0} \;.
\end{aligned}\end{equation}

Our final assumption is regarding the symmetry of the node dynamics~\eqref{eq:1-oscillator}. We assume that the node dynamics are homogeneous with respect to phase angles, i.e. there are no distinguished phase angles of $u$ and $z$ such that the dynamics can only depend on relative phase angles between the three-phase variables. This degree of freedom with respect to absolute phase angles translates to a symmetry under global phase shifts, extending the natural symmetry of the desired operating state~\eqref{eq:qss} to the whole phase space. As a global phase shift is equivalent to a time shift for the uniformly rotating operating state, this last assumption is essentially the requirement that in a quasi-steady state it doesn't matter when a perturbation hits of time invariance with respect to perturbations of the operating state, a highly desired property of control systems. Note, however, that this assumption is only valid when the transistor switching may be modeled as ideal, i.e. when the three-phase signals do not feature higher harmonics but are purely sinusoidal. This implies that the dynamics must be on a time scale where the switching may safely be neglected by averaging~\cite{erickson2001, chiniforoosh2010}, which is a standard assumption in power grid control design considering the switching frequencies are typically in the range of 2-20 kHz~\cite{schiffer2016}. 

\begin{assumption}[Symmetry]
\label{as:homogeneity}
The node dynamics is homogeneous with respect to phase angles, i.e. it possesses a U(1) symmetry defined by
\begin{equation}\label{eq:symmetry}\begin{aligned}
f^{u,z}&(x, e^{i\theta} z, e^{-i\theta} z^*, e^{i\theta} u, e^{-i\theta} u^*, e^{i\theta} j, e^{-i\theta} j^*) \\&= e^{i\theta} f^{u,z}(x, z, z^*, u, u^*, j, j^*) \;,\\
f^x&(x, e^{i\theta} z, e^{-i\theta} z^*, e^{i\theta} u, e^{-i\theta} u^*, e^{i\theta} j, e^{-i\theta} j^*) \\  &=   f^x(x, z, z^*, u, u^*, j, j^*) \;,
\end{aligned}\end{equation}
for any $\theta \in [0, 2\pi)$.
\end{assumption}

\subsubsection{Networked model}

When the general form \eqref{eq:1-oscillator} is combined with a model for the transmission lines providing the dynamics of the nodal currents, we obtain a connected model for the power grid. While the subsequent derivation of the normal form does not assume a particular model for the transmission lines, for the discussion we will make use of the very simple model of static currents, which we briefly want to introduce here. A more detailed discussion of dynamical current models is given in the appendix \ref{sec:appendix_line_dynamics}.

We define the network as a graph $\mathcal{G} = (\mathcal{V}, \mathcal{E})$ with the set of vertices $\mathcal{V} = \{1,\dots,N\}$, the set of edges $\mathcal{E} = \{1,\dots,M\}$, and its complex structure given by the incidence matrix $B \in \{-1,0,1\}^{N\times M}$. Additionally, for each edge we have its resistance $r_m \in \mathbb{R}_{\geq 0}$ and its inductance $\ell_m \in \mathbb{R}_{\geq 0}$.

Assuming that the line dynamics evolve much faster than the node dynamics, and that we are in the vicinity of the desired operating state with uniform frequency $\Omega_s$, the nodal currents may be approximated by its quasi-steady-state equations. In terms of the admittance matrix $Y := B~\mathrm{diag}\left(r_m + i\Omega^s \ell_m\right)^{-1} B^T$, these are simply given by
\begin{equation}\label{eq:ss_algebraic}
	j_n = \sum_{m=1}^N Y_{nm} u_m \;.
\end{equation}

It follows that we have the active and reactive power inputs/outputs, $p_n + iq_n := u_n j_n^*$, given by
\begin{equation}\label{eq:ss_algebraic_pq}
	p_n + iq_n = \sum_{m=1}^N Y_{nm}^* u_n u_m^* \;.
\end{equation}

Note that these equations, as well as the full dynamics from which they arise, are consistent with the symmetry of assumption~\ref{as:homogeneity}. This means that the full network model is invariant under global phase shifts. Quasi-steady states are again given by an orbit of the symmetry. However a quasi-steady state of the coupled model is not guaranteed to exist, unless the set points of the nodes are chosen in a manner compatible with \eqref{eq:ss_algebraic_pq}. If this is not the case, for example, if there is a power imbalance in the system, the network's quasi-steady state will deviate from the orbit specified by the set points, i.e. the desired operating state.

\subsection{Normal form of the node dynamics}

The system~\eqref{eq:1-oscillator} is highly general and captures most models for grid-forming components, as long as they do not explicitly feature higher level control layers, asymmetric phases, non-smooth features (like current limitation, as in~\cite{doerfler2019b} for example), algebraic equations without a closed-form solution, or model-free control (e.g. data-based approaches \cite{coulson2019data}). The central insight is that by exploiting the symmetry we can cast the model in a form that is given by an explicit dependence on the voltage $u$ that is tightly constraint by the symmetry, and a set of meaningful quadratic invariants. The latter uniquely specify different orbits of the symmetry~\eqref{eq:symmetry}. An orbit of the symmetry is a set of states related by phase shifts. Thus the quasi-steady states correspond to such orbits. Therefore, the invariants provide us with a sensible notion of the distance to the desired operating state (or any other specified quasi-steady state). We can then develop the dynamics order by order in these quadratic invariants. This provides us with a normal form for grid components with dynamics of the form of eqns.~\eqref{eq:1-oscillator} that is valid in the vicinity of the desired operating state.


\subsubsection{Derivation}

To derive the normal form we will make a temporary change of variables based on quadratic invariants of the symmetry~\eqref{eq:symmetry}. Recall the equations for the node dynamics
\begin{equation}\tag{\ref{eq:1-oscillator}}\begin{aligned}
	\dot{u} &= f^u(x, z, z^*, u, u^*, j, j^*) \;,\\
	\dot{z} &= f^z(x, z, z^*, u, u^*, j, j^*) \;,\\
	\dot{x} &= f^x(x, z, z^*, u, u^*, j, j^*) \;.
\end{aligned}\end{equation}

Since the symmetry is defined by a 1-dimensional $U(1)$ action on a $2L + 4$ dimensional space (disregarding the scalar variables $x$ for that matter), there are $2L + 3$ functionally independent invariants associated with it (see e.g. \cite[Theorem 2.17]{olver1993}). These invariants are sufficient to specify the orbits of the group action for an individual node and thus also its desired operating state. Therefore, we want to choose physically meaningful invariants, including, in particular, the quantities that are typically used to define the operation point of the power grid. While there is a certain freedom of choice involved, we propose the set of invariants comprising $u u^* =: \rho^2$, the voltage amplitude squared, $(u j^* + u^* j)/2 =: p$ and $(u j^* - u^* j)/(2i) =: q$, active and reactive power input/output, as well as $(u z^* + u^* z)/2 =:\psi$ and $(u z^* - u^* z)/(2i) =:\chi$, which may be interpreted as internal active and reactive power flows with respect to the terminal voltage. As the remaining variable that is needed to fully describe the node dynamics, we choose to keep the complex voltage $u$.

In this combination all invariants are real-valued, polynomial, and, together with the complex voltage $u$, do not require division by the current or any of the internal variables when used to express the original set of variables, i.e. the coordinate transformation is never singular away from the origin $u = u^* = 0$. Employing this new set of variables and defining $\xi := (x^T, \psi^T, \chi^T)^T \in \mathbb{R}^{K + 2L}$ yields the system
\begin{equation}\label{eq:transformed}\begin{aligned}
\dot{u} &= \tilde{f}^u(u, \xi, \rho^2, p, q)  \;,\\
\dot{\xi} &= \tilde{f}^\xi(u, \xi, \rho^2, p, q)  \;.
\end{aligned}\end{equation}
The explicit transformation is spelled out in Appendix~\ref{sec:appendix_transformation}. Note that instead of depending on $u$ and $u^*$ we have $u$ and $\rho^2 = u u^*$. Effectively we have expressed a non-holomorphic function of one complex variable through a function of one complex and one real variable that is holomorphic in the former. Also, note that we did not explicitly write down the remaining dynamical equations for $\rho^2$, $p$ and $q$, as we are ultimately interested in a normal form expressed in terms of $u$ and $j$ directly and will not thus need these equations in the end.

Now the symmetry conditions~\eqref{eq:symmetry} can be stated in terms of the infinitesimal generator of the group action~\cite{olver1993} as
\begin{equation}\label{eq:symmetry_infinitesimal_new}\begin{aligned}
u\frac{\partial \tilde{f}^u}{\partial u} = \tilde{f}^u \;,\quad 
\frac{\partial \tilde{f}^\xi}{\partial u} = \vec{0} \;.
\end{aligned}\end{equation}
These equations can be readily integrated and yield the node dynamics
\begin{equation*}\begin{aligned}
\dot{u} &= u g^u(\xi, \rho^2, p, q)  \;,\\
\dot{\xi} &= \phantom{u} g^\xi(\xi, \rho^2, p, q) \;,
\end{aligned}\end{equation*}
with some continuous nonlinear functions $g^u:~\mathbb{R}^{K + 2L + 3} \to \mathbb{C}$ and $g^\xi: \mathbb{R}^{K + 2L + 3} \to \mathbb{R}^{K + 2L}$. What we have thus achieved is a unified description of all possible node dynamics congruent with our modeling assumptions, that is based on the common symmetry of its quasi-steady state and preserves the physical relationship between current, voltage and power. As it turns out, we may describe the node dynamics as a complex oscillator that is augmented by some internal dynamics and is coupled to the other nodes in the network via active and reactive power. Expressing the system in this way now allows for the aforementioned Taylor expansion with a well-defined notion of closeness to some desired operating state.

For notational convenience we define the vector of instantaneous invariants $y(t) := (\xi(t)^T, \rho(t)^2, p(t), q(t))^T: \mathbb{R} \to \mathbb{R}^{K + 2L + 3}$ and a given constant vector $y_0 := (\xi^T_0, \rho^2_0, p_0, q_0)^T \in \mathbb{R}^{K + 2L + 3}$ with $\rho_0 > 0$ denoting the point around which we want to expand the node dynamics. While the latter can in principle be chosen arbitrarily, of most practical interest are the cases when $y_0$ represents either (i) a valid operating state considering the entire network, that is, the active and reactive power and the voltage amplitude are consistent with a power flow solution, or (ii) the set-points (or also educated guesses thereof if not explicitly available), which may or may not be consistent with a power flow solution. However, in both cases we would have $g^u(y_0) = i\Omega_s$ and $g^\xi(y_0) = \vec{0}$.

Denoting the deviation of $y$ from $y_0$ by $\delta y := y - y_0$, up to first order we have
\begin{equation}\label{eq:normal_form}\begin{aligned}
\frac{\dot{u}}{u} &= g^u(y_0) + (\delta y \cdot \nabla) g^u (y_0) + \mathcal{O}(\lVert \delta y \rVert^2) \\
&\simeq A^u + B^u \delta\xi + C^u \delta\rho^2 + G^u \delta p + H^u \delta q \;, \\[8pt]
\dot{\delta\xi} &= g^\xi(y_0) + (\delta y \cdot \nabla) g^\xi (y_0) + \mathcal{O}(\lVert \delta y \rVert^2) \\[4pt]
&\simeq A^\xi + B^\xi \delta\xi + C^\xi \delta\rho^2 + G^\xi \delta p + H^\xi \delta q  \;,
\end{aligned}\end{equation}
with the respective coefficients $C^u, G^u, H^u \in \mathbb{C}$, $B^u \in \mathbb{C}^{1\times (K + 2L)}$, $C^\xi, G^\xi, H^\xi \in \mathbb{R}^{K + 2L}$, $B^\xi \in \mathbb{R}^{(K+2L)\times (K+2L)}$, and $A^{u,\xi} := g^{u,\xi}(y_0)$. 

\subsubsection*{Normal form of the node dynamics}
Having carried out the expansion, we may now return to using the dynamical variables $u$, $u^*$ and $j$, $j^*$ so that our model may be easily connected by providing equations for the currents flowing through the network according to some model for the transmission lines. We thus arrive at the normal form
\begin{equation}\label{eq:normal_form_uj}\begin{aligned} 
	&\delta p + i\delta q = u j^* - (p_0 + i q_0)\\
	&\delta\rho^2 = u u^* - \rho_{0}^2 \;,\\
    &\frac{\dot{u}}{u} \simeq A^u + B^u \delta \xi + C^u \delta\rho^2 + G^u \delta p + H^u \delta q \;, \\[8pt]
	&\dot{\delta\xi} \simeq A^\xi + B^\xi \delta \xi + C^\xi \delta\rho^2 + G^\xi \delta p + H^\xi \delta q\;.
\end{aligned}\end{equation}

Note that the back-transformation does not change the quality of the approximation, this normal form is still accurate up to terms of order $\lVert \delta y \rVert^2$. Moreover, since we are dealing with asymptotically stable systems, we expect the error to be bounded and, in most practical cases, quickly decrease over time as long as the trajectories remain within the basins of attraction of both the original system and its normal form.

The rest of this paper will explore the implications and properties of this normal form.


\begin{figure*}[!th]
\begin{subfigure}{0.32\textwidth}
\includegraphics[width=\textwidth]{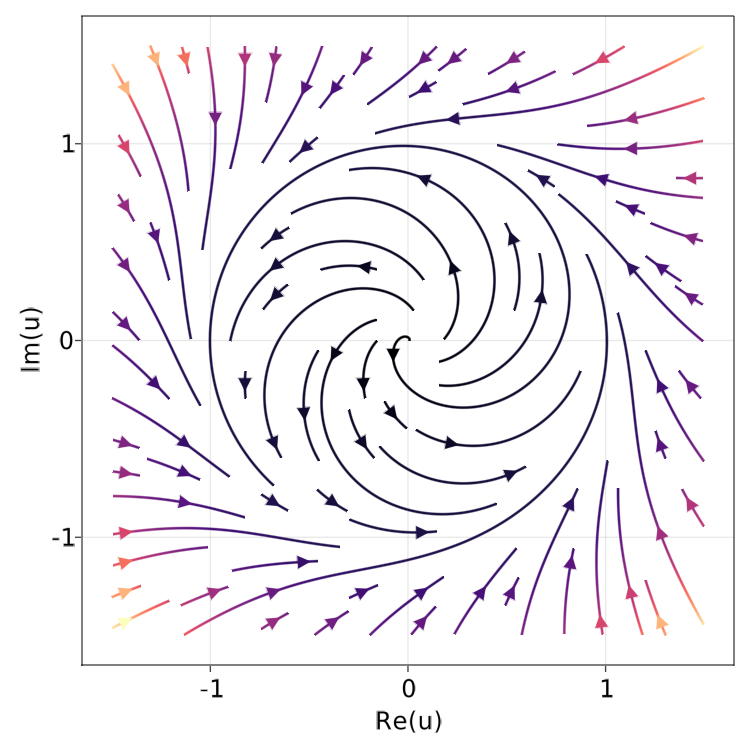}
\caption{}
\end{subfigure}
\begin{subfigure}{0.32\textwidth}
\includegraphics[width=\textwidth]{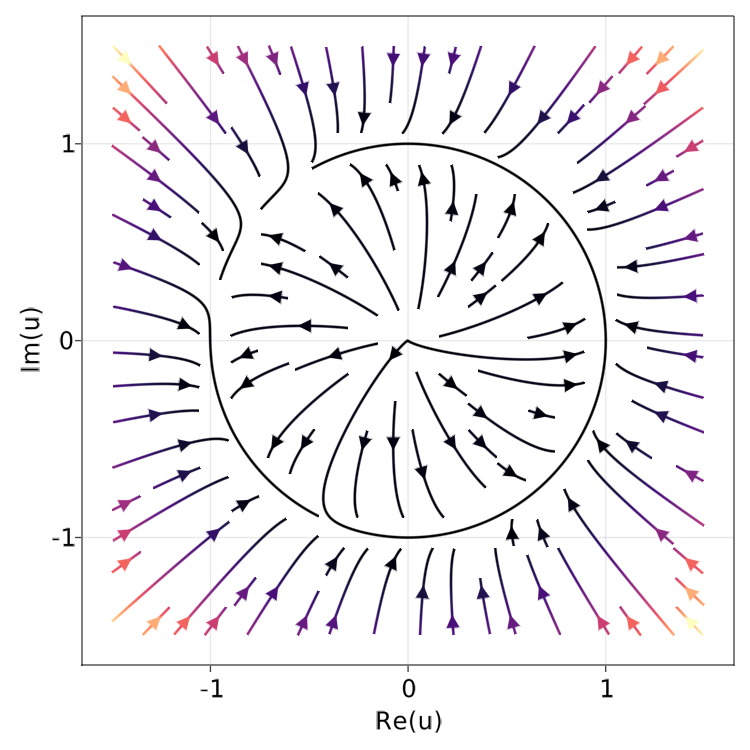}
\caption{}
\end{subfigure}
\begin{subfigure}{0.32\textwidth}
\includegraphics[width=\textwidth]{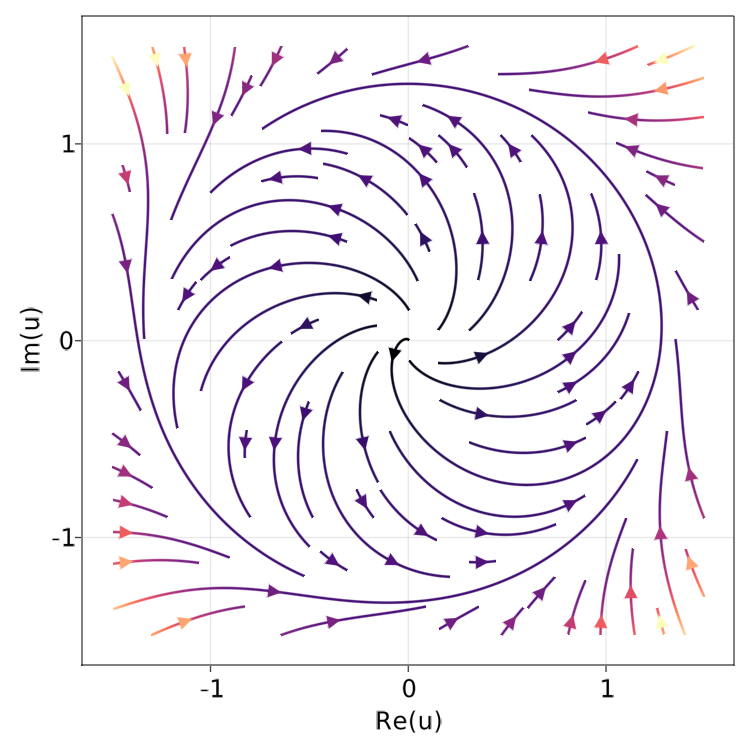}
\caption{}
\end{subfigure}
\caption[]{Streamplot example for the normal form model without any internal variables. The parameters are $A^u = 1+i$, $C^u = -1$, $G^u = i$ and $H^u = 1$. For this parametrization the trajectories converge to a stable limit cycle. The dynamical behavior changes for deviations of the power: (a) no power deviation ($\delta p = \delta q = 0$), (b) an active power deviation ($\delta p = -0.9 $, $\delta q = 0$) and (c) a reactive power deviation ($\delta p = 0$, $\delta q = 0.7$). It can be seen that for this choice of parameters, a change in active power changes the angular velocity (i.e. the frequency), whereas a change in reactive power changes the amplitude of the limit cycle (i.e. the voltage amplitude).}
\label{fig:streamplots}
\end{figure*}

\subsubsection{Interpretation}


To connect the equations \eqref{eq:normal_form_uj} to more familiar models for power grids it is insightful to consider them from the point of view of phase-amplitude coupling in power grids. To this end, consider the complex voltage in terms of phase and the logarithm of the amplitude $\sigma$, i.e. $u = e^{\sigma + i\phi}$ and
\begin{equation}\label{eq:phase-amplitude}
    \frac{\dot{u}}{u} = \dot \sigma + i \dot \phi \;.
\end{equation}
As $\delta \xi$, $\delta\rho^2$, $\delta p$ and $\delta q$ are all real-valued, the real and imaginary part of $A^u$, $B^u$, $C^u$, $G^u$, and $H^u$ respectively control the influence of the corresponding terms on amplitude and phase dynamics. For example, in the absence of internal variables the explicit impact of amplitude deviations on phase dynamics can be read off immediately from the imaginary part of $C^u$. However, internal dynamics can mix phase and amplitude reactions in more subtle ways.



The form \eqref{eq:normal_form_uj} expands the internal dynamics to linear order in the quadratic invariants. The approximate internal dynamics is a linear multi-input-multi-output system (MIMO) with two outputs, given by the real and imaginary part of the right hand side of $\dot{u}/u$, and three non-constant inputs, given by $\delta \rho^2$, $\delta p$ and $\delta q$. The explicit form is given in Appendix~\ref{sec:appendix_mimo}.

This opens the door to introducing methods from the analysis of linear time invariant systems, such as model order reduction, to the study of power grid models in a systematic fashion. One way in which we will already touch upon this possibility later in the paper is by using system identification techniques to fit a model of a fixed complexity to measurement data of a real inverter. This provides semi-black box models for components (see section~\ref{sec:tecnalia}).


We also want to remark that, while we chose to work with the complex voltages (as it makes the expression for active and reactive power particularly simple, is in line with the recently developed concepts of virtual oscillator control~\cite{johnson2016, johnson2017, doerfler2018}, and bears resemblance to an already established dynamics as discussed in section~\ref{sec:stuart-landau}),  it is not mandatory to work with these coordinates. If desired, it is also possible to carry out our approach in terms of $u_\alpha := \Re(u)$ and $u_\beta := \Im(u)$ or amplitude and phase directly, i.e. $u =: \rho e^{i\phi}$.



\subsection{Relation to other models}\label{sec:examples}

By approximating the fully nonlinear system~\eqref{eq:1-oscillator} with our normal form~\eqref{eq:normal_form_uj}, we have, in principle, reduced all qualitative differences between various concrete models to quantitative differences in their respective coefficients. Thereby, we can distinguish certain classes of models by the coefficients being non-zero. In this section, we introduce a few simple examples of different classes of power-grid models and their corresponding normal form, provide working parameter values, and discuss how the normal form relates to the well-known Stuart-Landau oscillator.

\subsubsection{Examples}

To understand the relationship of the normal form to established low-dimensional models of grid-forming power grid components we will provide the normal form approximation of a variety of them, and demonstrate which coefficients occur in various classes of models. These examples will also be used in the numerical experiments for a heterogeneous network in Section \ref{sec:numerics-network}.


We begin by giving a concrete implementation of the abstract derivation of the preceding section for the droop-controlled inverter model introduced in \cite{schiffer2016} (to be self contained we give the equations in the appendix, eqns.~\eqref{eq:schiffer}). In this model the dynamical equation for the frequency, the only internal variable, is already linear with respect to the invariants. As the dynamical equations for the voltage are formulated in terms of amplitude and phase, i.e. $\rho = \lvert u \rvert$ and $\phi = \arg u$, we can make use of the relationship 
\begin{equation*}
    \dot{u} = \left( \frac{\dot{\rho}}{\rho} + i\dot{\phi} \right) u \;,
\end{equation*}
to translate them into the complex form. This yields
\begin{align*}
    \frac{\dot{u}}{u} &= \frac{1}{\tau_{p} \rho} \left( -\rho + \rho^d - k_{q} (q - q^d)\right) + i\omega \\
    &= i\omega + \frac{\rho^d}{\tau_{p} \rho} -  \frac{1}{\tau_{p}} - \frac{k_{q}}{\tau_{p} \rho}(q - q^d)  \;.
\end{align*}
Now carrying out the expansion for $y_{0} = (\omega^d, (\rho^d)^2, p^d, q^d)^T$, as prescribed by eqns.~\eqref{eq:normal_form}, we arrive at
\begin{equation}\label{eq:schiffer_normal}\begin{aligned}
    \frac{\dot{u}}{u} &\simeq i\delta\omega - \frac{1}{2 \tau_{p} \rho_{0}^2} \delta\rho^2 - \frac{k_{q}}{\tau_{p} \rho_{0}} \delta q \;,\\
    \dot{\delta\omega} &= -\frac{1}{\tau_{p}} \delta\omega - \frac{k_{p}}{\tau_{p}} \delta p \;.
\end{aligned}\end{equation}

The table of normal form coefficients in terms of the original parameters is given by the following:
\begin{center}
\begin{tabular}{c|c|c|c|c|c}
    & $A$ & $B$ & $C$ & $G$ & $H$ \\
    \hline
    $u$ & $i\Omega_s$ & $i$ & $- \frac{1}{2 \tau_{p} \rho_{0}^2}$ & $0$ & $- \frac{k_{q}}{\tau_{p} \rho_{0}}$     \\
    \hline
    $\omega$ & $0$ & $-\frac{1}{\tau_{p}}$  & $0$ & $- \frac{k_{p}}{\tau_{p}}$ & $0$ \\
\end{tabular}
\end{center}

In a similar fashion, we can consider other classes that result from well-known models.

\begin{example}[Pure phase oscillators]
Pure phase oscillators are oscillators without any amplitude or internal dynamics. As discussed above in light of equation \eqref{eq:phase-amplitude}, the imaginary part of the coefficients provides the phase dynamics, the real part the amplitude dynamics. Pure phase oscillators thus yield the normal form
\begin{equation*}
	\frac{\dot{u}}{u} \simeq A^u + G^u \delta p + H^u \delta q
\end{equation*}
with $A^u,G^u,H^u \in i\mathbb{R}$. The canonical example from this class is the well-known Kuramoto model~\cite{rodrigues2016, simpson2013synchronization}, which further has $H^u = 0$ and assumes the nodes are coupled by purely inductive transmission lines, i.e. $\Re(Y) = 0$.
\end{example}

\begin{example}[Phase-frequency oscillators]
The ubiquitous swing equation~\cite{machowski2020} (or its nonlinear variant~\cite{monshizadeh2016}) falls into the class of phase-frequency oscillators. We still have no amplitude dynamics, but now allow for an internal variable: the frequency of the oscillator. The normal form of this type of oscillator is given by
\begin{equation*}\label{eq:swing_eq_app}\begin{aligned}
	\frac{\dot{u}}{u} &= A^u + i\delta\omega \;,\\
	\delta\dot{\omega} &\simeq A^\omega + B^\omega \delta\omega + G^\omega \delta p + H^\omega \delta q \;,
\end{aligned}\end{equation*}
with $A^u \in i\mathbb{R}$. As for the Kuramoto model, the standard swing equation further has $H^\omega = 0$ and typically assumes coupling with $\Re(Y) = 0$.
\end{example}

\begin{example}[Phase-amplitude oscillators]
The quite recent development of virtual oscillator control~\cite{johnson2016, johnson2017, doerfler2018} represents phase-amplitude oscillators, i.e. there is now amplitude dynamics but no internal dynamics. The corresponding normal form is given by
\begin{equation}\label{eq:dVOC_app}\begin{aligned}
	\frac{\dot{u}}{u} &\simeq A^u + C^u \delta\rho^2 + G^u \delta p + H^u \delta q \;.
\end{aligned}\end{equation}
\end{example}

\begin{example}[Phase-amplitude-frequency oscillators]
Synchronous machines can also be cast into normal form. There are both phase and amplitude dynamics as well as internal dynamics. Considering third order models~\cite{schmietendorf2013}, there is only the frequency as internal variable, and the normal form for this type of model reads as
\begin{equation}\label{eq:third_order_app}\begin{aligned}
	\frac{\dot{u}}{u} &\simeq A^u + B^u \delta\omega + C^u \delta\rho^2 + G^u \delta p + H^u \delta q \;,\\
	\delta\dot{\omega} &\simeq A^\omega + B^\omega \delta\omega + C^\omega \delta\rho^2 + G^\omega \delta p + H^\omega \delta q \;,
\end{aligned}\end{equation}
with $C^u, G^u, H^u \in \mathbb{R}$ if $\omega$ should be the true frequency of the nodal voltage. As we see from eqns. \eqref{eq:schiffer_normal}, the droop-controlled inverter of \cite{schiffer2014, johnson2017} also falls into this class.
\end{example}

\subsubsection{Normalized parameters values}

If one wants to work with our normal form, eqns.~\eqref{eq:normal_form_uj}, as an abstract model for grid-forming components detached from any concrete models, there is the question of reasonable parameter values from which to start exploring the parameter space and its dynamic features. To spare the reader the tedious work of trial and error, we thus want to give a starting point here. As the dynamics of the system will depend highly on the given network structure and node types, we here consider the simple case of two identical grid-forming components connected by an RL transmission line. We work in dimensionless units such that $\lvert r + i\Omega_s \ell \rvert = 1$ and define $\tan\kappa := \Omega_s\ell/r$ to account for the ratio of inductivity and resistivity of the transmission line (more details on this in section~\ref{sec:linear_stability}). Prescribing an operating state with $\rho_{0,1} = \rho_{0,2} = 1$ (in dimensionless units) and a relative phase angle of $\lvert \Delta\phi_0 \rvert < \cos^{-1}(1/2)$, the following sets of parameters yield convergence to the operating state on a time scale of 5-10 seconds:
\begin{itemize}
    \item Phase-amplitude oscillators (eqn.~\eqref{eq:dVOC_app})
    \begin{center}
        \begin{tabular}{c|c|c|c|c}
        & $A$ & $C$ & $G$ & $H$ \\
        \hline
        $u$ & $i\Omega_s$ & $-1.5$ & $-0.8 (\cos\kappa + i\sin\kappa)$ & $-0.8 (\sin\kappa - i\cos\kappa)$     \\
        \end{tabular}
    \end{center}
    \item Phase-amplitude-frequency oscillators (eqns.~\eqref{eq:third_order_app})
    \begin{center}
    \begin{tabular}{c|c|c|c|c|c}
        & $A$ & $B$ & $C$ & $G$ & $H$ \\
        \hline
        $u$ & $i\Omega_s$ & $i$ & $-1.5$ & $-0.8\cos\kappa$ & $-0.8\sin\kappa$     \\
        \hline
        $\omega$ & $0$ & $-1.5$ & $0$ & $0.8\cos\kappa$ & $-0.8\sin\kappa$ \\
    \end{tabular}
    \end{center}
\end{itemize}
For more realistic sets of parameters we refer to the simulations of section~\ref{sec:validation}, where we numerically compare the normal form to concrete models. The parameters used there are of similar order of magnitude however.

\subsubsection{Connection to the Stuart-Landau oscillator}\label{sec:stuart-landau}

The normal form derived above is closely related to the classical model known as the Stuart-Landau oscillator~\cite{kuramoto1984}. Key to this connection is that the invariants we chose to represent the deviation from the limit cycle, in particular the voltage amplitude squared, are polynomial with respect to $u$ and $u^*$.

Using the quasi-steady state approximation for the current eqn.~\eqref{eq:ss_algebraic}, and the normal form for amplitude-phase dynamics, eqns.~\eqref{eq:normal_form_uj}, yields the networked system
\begin{equation}\label{eq:normal_form_SL}\begin{aligned}
\dot{u}_n \simeq & ~\left( \tilde{A}_n + C_n \lvert u_n \rvert^2 \right) u_n\\
&~+ \sum_{m=1}^{N} \left( {K}^+_n Y_{nm}^* u_n^2 u_m^* + {K}^-_n Y_{nm} \lvert u_n \rvert^2 u_m \right) \;, 
\end{aligned}\end{equation} 
where we have absorbed the expansion points into the coefficients, i.e.
\begin{equation*}\begin{aligned}
    \tilde{A}_n &:= A^{u}_n - C^{u}_n \rho^2_{0,n} - G^{u}_n p_{0,n} - H^{u}_n q_{0,n} \;,\\
	{K}^\pm_n &:= \frac{1}{2} (G^{u}_n \pm i H^{u}_n) \;.\\
\end{aligned}\end{equation*}

It can be seen that the dynamics of the complex voltages $u_n$ is that of Stuart-Landau oscillators with a particular non-linear coupling. Since the admittance matrix $Y$ is given as a linear combination of Laplacian matrices, the coupling between the individual oscillators may be interpreted as diffusive, albeit with a state-dependent diffusion matrix and also involving the complex conjugate voltages of the connected nodes.

This resemblance is of course no coincidence as the Hopf bifurcation (from which the Stuart-Landau oscillator results as the corresponding normal form~\cite{kuznetsov1998}) prescribes the same U(1) symmetry for the emerging limit cycle in the vicinity of the bifurcation point. Furthermore, the nonlinearity of the amplitude squared arises naturally in the context of a Taylor expansion. This close relationship opens the door to applying methods from the study of coupled Stuart-Landau oscillators, as for example~\cite{panteley2015, panteley2016, roehm2018}, to the dynamics the power grids.

\subsection{Validation and probabilistic stability}\label{sec:validation}

\begin{figure*}[th!]
\includegraphics[width=0.48\textwidth]{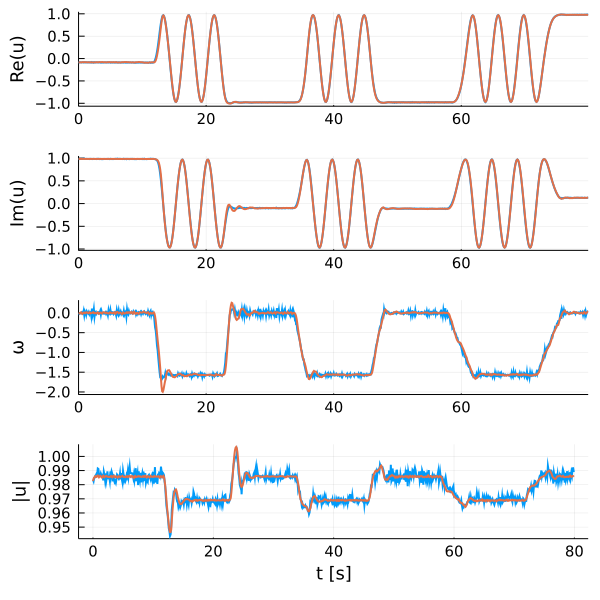}
\includegraphics[width=0.48\textwidth]{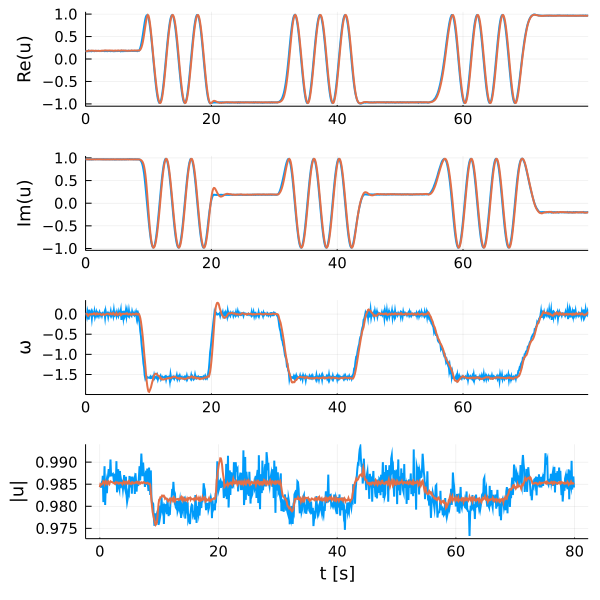}

\caption[]{Left: Model fit to data, one internal variable. Right, validation of the fitted parameters against different test run. Blue: measurement data. Orange: output of the normal form.}
\label{fig:fit_to_data}
\end{figure*}

While the normal form is guaranteed to be a valid approximation in a small neighborhood of the desired operating state, it is not a priori clear whether it can successfully approximate the behavior of real systems under realistic perturbation. Further, if the quasi-steady state of the network starts to deviate from the desired operating state of the individual units we might be stretching the validity of the normal form even further. This section will show first evidence that real systems (section~\ref{sec:tecnalia}) and large perturbations (sections~\ref{sec:numerics-inf-bus}) can be accurately captured by the normal form. Further we show that for an adapted standard IEEE test network, the normal form approximation correctly captures both, persistent deviations from the desired operating state as well as probabilistic stability properties in section~\ref{sec:numerics-network}.

All code is available and open source in an accompanying github repository/zeonodo archive.



\subsubsection{Lab experiment} \label{sec:tecnalia}
We begin with an empirical test of our model using measurement data of a grid-forming inverter with an elaborate control scheme, devised and built at TECNALIA labs \cite{planas2013design}. The data has originally been gathered to validate numerical simulation tools \cite{plietzsch2021powerdynamics}.


We use the normal form as a semi-black box model that we fit to this measurement data. Note that while a detailed model of the inverter may include many inner control loops, all of which need to be modeled correctly to reproduce the measurements, the semi-black box model can be chosen to be much simpler. In fact, we will use a single internal variable, thus obtaining an effective model of reduced order (dimensionality) for the inverter. Details on the measurement setup and the fitting procedure can be found in the \emph{Materials and Methods} section.



Figure~\ref{fig:fit_to_data} depicts the results. On the left, we see the measurement data against the trajectories of the optimized normal form with the input that has been used for the optimization. On the right, we show the trajectories for an input from a different measurement while using the same set of parameters. We see that the normal form is able to capture most of the dynamical behavior of the grid-forming inverter very accurately, only showing overshoots during the sudden shifts in frequency. This is to be expected, however, as we have used a model of quite low dimensionality.

\subsubsection{Simulations - infinite bus}\label{sec:numerics-inf-bus}

\begin{figure*}[!th]
\begin{subfigure}{0.34\textwidth}
\includegraphics[width=\textwidth]{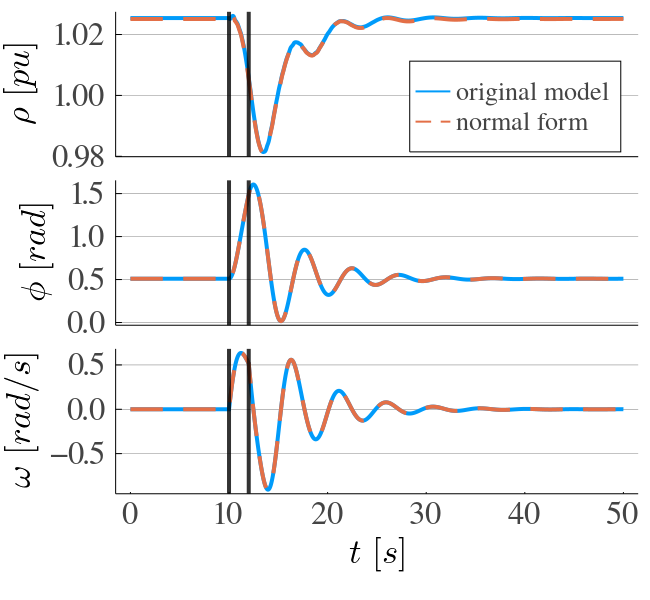}
\caption{}
\end{subfigure}
\begin{subfigure}{0.3\textwidth}
\includegraphics[width=\textwidth]{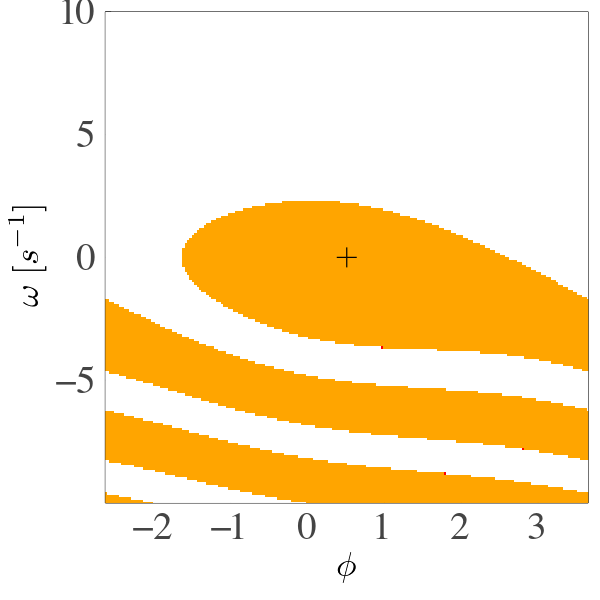}
\caption{}
\end{subfigure}
\begin{subfigure}{0.3\textwidth}
\includegraphics[width=\textwidth]{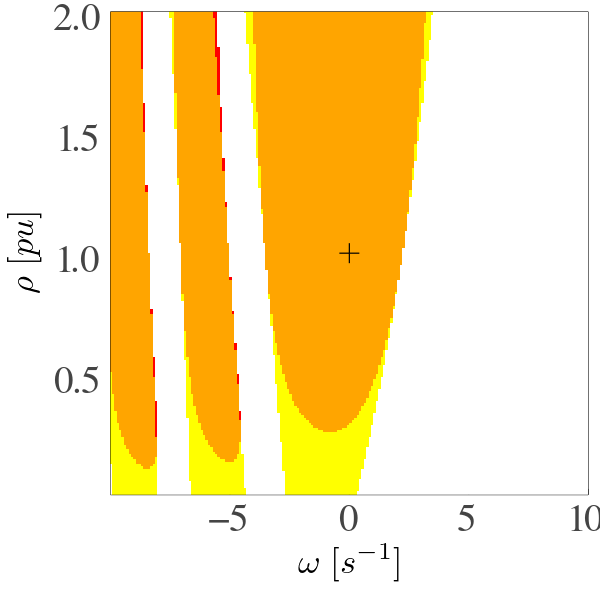}
\caption{}
\end{subfigure}
\\
\begin{subfigure}{0.34\textwidth}
\includegraphics[width=\textwidth]{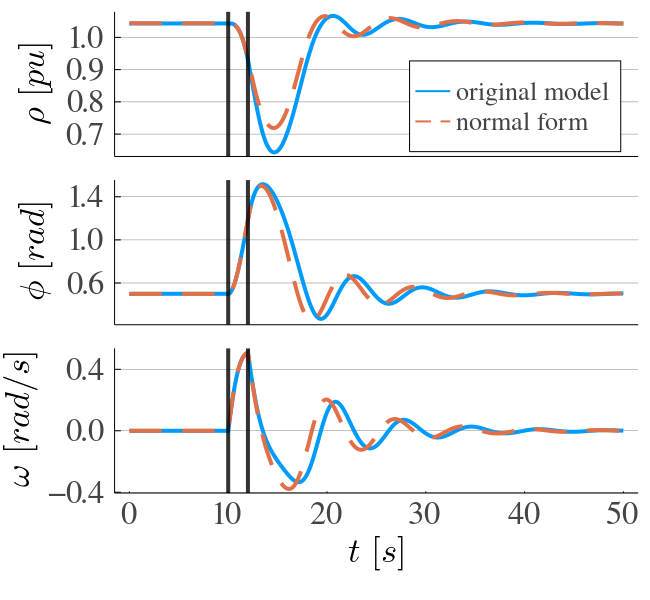}
\caption{}
\end{subfigure}
\begin{subfigure}{0.3\textwidth}
\includegraphics[width=\textwidth]{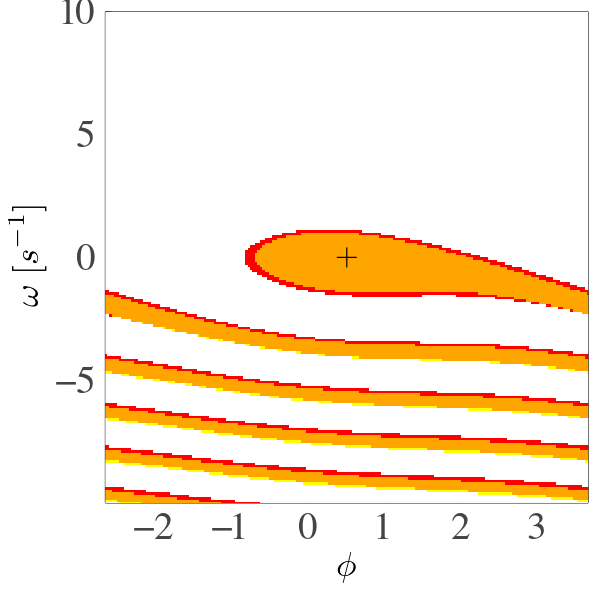}
\caption{}
\end{subfigure}
\begin{subfigure}{0.3\textwidth}
\includegraphics[width=\textwidth]{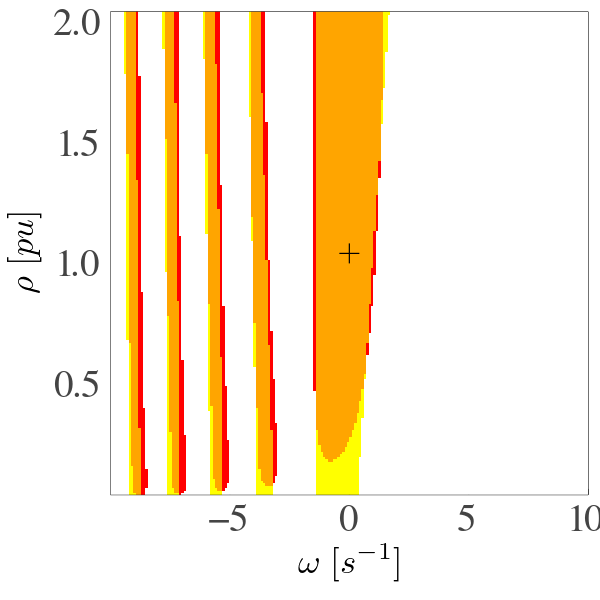}
\caption{}
\end{subfigure}
\caption[]{Results for the infinite bus scenario. Trajectories for (a) inverter, (d) synchronous machine (the vertical bars signify the beginning and end of the perturbation at $t=10s$ and $t=12s$). Basin cross-sections for (b, c) inverter, (e, f) synchronous machine with four possible cases for each state: points inside the basins of both the original model and its normal form (orange), only inside the original model's basin (yellow), only inside the normal form's basin (red), and outside both basins (white). The crosses depict the actual operating state $(\phi^o, \omega^o, \rho^o)$ of the full models.}
\label{fig:infinite_busbar_results}
\end{figure*}

We now turn to simulation studies of the normal form approximation in an infinite bus bar setting, as already considered in section \ref{sec:linear_stability}. We will consider droop-controlled inverters of \citet{schiffer2014} (see eqns.~\eqref{eq:schiffer}) and third order models of synchronous machines as used by \citet{schmietendorf2013} (see eqns.~\eqref{eq:schmietendorf}).

Figures~\ref{fig:infinite_busbar_results}a and~\ref{fig:infinite_busbar_results}d show the resulting trajectories for a large power perturbation during which the desired power input at the node is doubled for two seconds. We see that the qualitative agreement of the trajectories is excellent in both cases. In fact the trajectories almost completely match for the droop-controlled inverter. For the synchronous machine, for which the trajectory drops to extremely low voltage levels, deviations are more noticeable.

To explore the reaction to a large perturbation more systematically, we consider various slices of the systems phase space. That is, given the operating state specified in amplitude-angle coordinates by $(\phi^o, \omega^o, \rho^o)$, we consider trajectories starting from coordinates of the form $(\phi, \omega, \rho^o)$ and $(\phi^o, \omega, \rho)$. We then ask whether the system returns to the operating state from these initial conditions, i.e. we consider slices of its basin of attraction. The model's basin is depicted in yellow, the basin for the normal form in red, and their overlap in orange.

Figure~\ref{fig:infinite_busbar_results}b shows that for the droop-controlled inverter the basins show full agreement. In figure~\ref{fig:infinite_busbar_results}c we see some deviations at low voltage amplitudes very far from the operating state. This is to be expected as the normal form approximation involved expanding the voltage dynamics around the desired operating state. In this case the normal form underestimates the stability region of the system. In the Appendix~\ref{sec:numerics-network} we see that this is not always the case.

For the third order model we have similar results in Figures~\ref{fig:infinite_busbar_results}e and~\ref{fig:infinite_busbar_results}f, with a slight overestimation of stability in some areas. The qualitative features of the original model are fully reproduced by the normal form though.

The details of the numerical set up are given in Appendix~\ref{sec:numerics_appendix}.

\subsubsection{IEEE-14 Bus system}\label{sec:numerics-network}

To address the question whether the normal form can also capture the complex interactions between different components, as well as persistent deviations in a realistic power grids, we turn to the IEEE 14-bus test system~\cite{canizares2003modeling}. We adapt this test system by placing various grid-forming component at the nodes, using synchronous machines, different types of droop control, and dVOCs. A detailed set up of the simulation is given in the appendix~\ref{sec:numerics_appendix}

\begin{figure}[!ht]
\includegraphics[width=0.95\columnwidth]{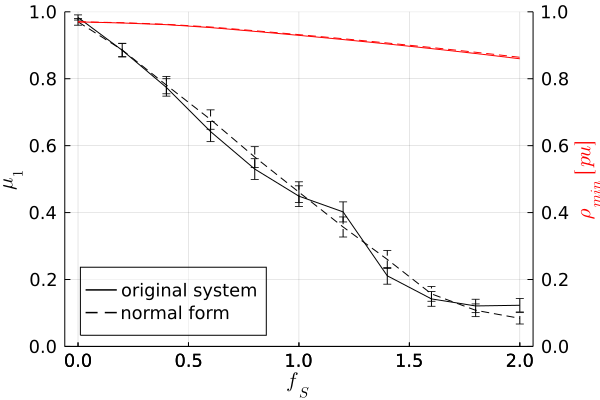}

\caption[]{IEEE 14-bus test system~\cite{canizares2003modeling}: The system is studied for a range of power demands, all active and reactive power set points at the inverter nodes are scaled by a common factor $f_s$ between $0$ and $2$. In red we plot the minimal voltage amplitude $\rho_{min} = \min_n \rho_n^o$ in the resulting operating state, the solid line is the original model, the dashed line the normal form. In black we give the single node basin stability of node 1.}
\label{fig:IEEE14_res}
\end{figure}

We then consider the normal form for all of them given their design set-points, ignoring deviations that result in the actual networks operating state due to imbalances and losses. We study the system at a range of power flows by scaling the active and reactive power at the inverter nodes by a common factor $f_s$ between $0$ and $2$. This leads to a variety of operating states which include some deviation from the desired operating points of the inverters. Figure~\ref{fig:IEEE14_res} shows in red the minimal voltage amplitude that occurs in the network for these operating states, for both the original and the normal form model. We see that the normal form is capable of describing this behavior very accurately. In black we show the single-node basin stability \cite{menck2013basin, menck2014dead} of node 1, i.e. the probability that a random large perturbation at node 1 destabilizes the system. Again, the results for the normal form and the full model agree within the uncertainty bounds.

This demonstrates that the normal form, by taking a "network first" approach to modeling and keeping the power flow equations fully accurate, is capable of capturing sophisticated properties of complex highly heterogeneous power grids.

\section*{Discussion}

This paper introduces a normal form for grid forming power grid components. This form is arrived at by using symmetry arguments to restrict the functional form and expanding order by order in physically meaningful quadratic non-linearities, concretely the power and voltage mismatch at the nodes. At lowest order the normal form is parametrized by a linear time invariant system with three time varying inputs and two outputs. The main factor in the complexity of the normal form is the number of internal states of this linear system.

The normal form can be derived from more detailed analytic models, for which we gave detailed examples, but it is also possible to directly infer it from experimental measurements. In the latter case, we can fix the number of internal states a priori to obtain an empirically best model of the system at a given complexity. We give a first proof of concept of this approach by fitting the data of a grid forming inverter built at TECNALIA to a low complexity normal form with one internal variable.

A more systematic exploration of this approach will require adapting tools from system identification to this context, especially for dealing with noise in the measurements.

We saw in numerical experiments that the normal form is capable of describing the non-linear behavior of the power grid in the vicinity of the desired operating states. This was explored for both, single machines at an infinite bus and a highly challenging heterogeneous network of diverse grid forming actors. While the quality of the lowest order approximation for a single node is relatively clear from the derivation and the numerical experiments, a more thorough understanding of the limits of the approximation when we consider a whole network of oscillators will require more work.

Besides being of interest in itself, the normal form presented here also provides a starting point for the transdisciplinary study of realistic models of future power grids \cite{brummitt2013transdisciplinary}.

The form closely resembles Stuart-Landau oscillators, thus opening the door to adapting a large body of dynamical systems research to the study of future power grids. As it is based on very general principles and physically meaningful variables all relevant dynamical aspects of the dynamical actors can be described by it. This opens up the possibility to transport results from the theoretical research on control and complex systems aspects of power grids, often based on highly conceptual phase models \cite{auer2016impact}, to models that are accurate with regard to the real power grid.

The normal form also opens up further novel research avenues. For example, it can serve not just as a model for concrete systems but as a specification for the behavior of future designs. The study of the linear stability of the normal form can be considered a first proof of concept in this direction.

While this work focuses on grid forming nodes, we expect the approach to be fruitful more broadly. By choosing different invariants, and different variables, it is possible to arrive at normal forms that will be suited to other classes of grid actors. Non-smooth behavior might also be modeled as switching between different normal forms.

Finally, the mathematical approach taken is highly general. Whereas ordinary phase reduction approaches require a small coupling assumption, here we were able to work from the assumption that coupling and oscillation are well adapted to each other. It is rare for oscillating systems to exist and develop in isolation, and we posit that such an approach is likely to be fruitful in other fields of complex systems science, synchronization and oscillator networks.

\section*{Materials and Methods}


For the empirical validation of our approach we use measurement data of a grid-forming inverter devised and built at TECNALIA labs. The inverter control design basically consists of conventional droop control with a low pass filter for the measured power output which emulates inertia. This basic design is similar to eqns.~\eqref{eq:schiffer}, but further includes additional filters for voltage and frequency measurements, as well as a virtual impedance~\cite{planas2013design}.\\

Besides the inverter, the lab setup contains an AC power source and an Ohmic load. The load is connected to the power source by an emulated line containing a series of inductances and resistances. The inverter is connected to the load by a transformer. More details on the setup and the parameterization of the components can be found in the technical report~\cite{VALERIA}. In this particular test case, we varied the voltage angle frequency at the power source and measured the voltages and currents for two of the three phases directly at the inverter. Assuming the three phases to be balanced, we can thus directly calculate the complex nodal voltage and current ($u$ and $j$), and further the active and reactive power output ($p$ and $q$), as well as the nodal voltage amplitude $\rho$. The frequency can be determined by numerical differentiation of the voltage phase angle.\\

We take the normal form with the voltage angle frequency as the only internal variable and fit to this measurement data. To reliably fit such models it will be necessary to properly adapt system identification techniques to this setting. To obtain a first proof of concept, we instead opted for a straightforward two step approach with generic tools. First, we performed a linear regression for obtaining rough parameter estimates, then we fine tuned these using scientific machine learning tools. For the linear regression, we first numerically calculate the derivative of the complex voltage and the frequency to obtain the left hand side of the differential equation \eqref{eq:third_order_app} and subsequently get an estimate for the parameters using the method of least squares. For the fine tuning, we use the current signal as a data-driven input for a dynamical simulation of the normal form model using the DifferentialEquations.jl package~\cite{rackauckas2017differentialequations} and optimize the least square fit of the trajectories with stochastic gradient decent using the DiffEqFlux.jl package~\cite{rackauckas2019diffeqflux}.

\section*{Code availability}

All code to reproduce the results and figures of this paper is available at the DOI \url{https://doi.org/10.5281/zenodo.4881898} or at the github repository \url{https://github.com/PIK-ICoNe/NormalFormPaper}.

\begin{acknowledgments}
We would like to thank Meng Zhan and Sebastian Liemann for detailed comments on a draft of this manuscript.

The authors acknowledge the support of BMBF, CoNDyNet2 FK. 03EK3055A.
This work was funded by the Deutsche Forschungsgemeinschaft (DFG, German Research Foundation) – KU 837/39-1 / RA 516/13-1 \& HE 6698/4-1.
All authors gratefully acknowledge the European Regional Development Fund (ERDF), the German Federal Ministry of Education and Research and the Land Brandenburg for supporting this project by providing resources on the high performance computer system at the Potsdam Institute for Climate Impact Research.
\end{acknowledgments}

\bibliography{references.bib}

\begin{thebibliography}{43}%
\makeatletter
\providecommand \@ifxundefined [1]{%
 \@ifx{#1\undefined}
}%
\providecommand \@ifnum [1]{%
 \ifnum #1\expandafter \@firstoftwo
 \else \expandafter \@secondoftwo
 \fi
}%
\providecommand \@ifx [1]{%
 \ifx #1\expandafter \@firstoftwo
 \else \expandafter \@secondoftwo
 \fi
}%
\providecommand \natexlab [1]{#1}%
\providecommand \enquote  [1]{``#1''}%
\providecommand \bibnamefont  [1]{#1}%
\providecommand \bibfnamefont [1]{#1}%
\providecommand \citenamefont [1]{#1}%
\providecommand \href@noop [0]{\@secondoftwo}%
\providecommand \href [0]{\begingroup \@sanitize@url \@href}%
\providecommand \@href[1]{\@@startlink{#1}\@@href}%
\providecommand \@@href[1]{\endgroup#1\@@endlink}%
\providecommand \@sanitize@url [0]{\catcode `\\12\catcode `\$12\catcode
  `\&12\catcode `\#12\catcode `\^12\catcode `\_12\catcode `\%12\relax}%
\providecommand \@@startlink[1]{}%
\providecommand \@@endlink[0]{}%
\providecommand \url  [0]{\begingroup\@sanitize@url \@url }%
\providecommand \@url [1]{\endgroup\@href {#1}{\urlprefix }}%
\providecommand \urlprefix  [0]{URL }%
\providecommand \Eprint [0]{\href }%
\providecommand \doibase [0]{https://doi.org/}%
\providecommand \selectlanguage [0]{\@gobble}%
\providecommand \bibinfo  [0]{\@secondoftwo}%
\providecommand \bibfield  [0]{\@secondoftwo}%
\providecommand \translation [1]{[#1]}%
\providecommand \BibitemOpen [0]{}%
\providecommand \bibitemStop [0]{}%
\providecommand \bibitemNoStop [0]{.\EOS\space}%
\providecommand \EOS [0]{\spacefactor3000\relax}%
\providecommand \BibitemShut  [1]{\csname bibitem#1\endcsname}%
\let\auto@bib@innerbib\@empty
\bibitem [{\citenamefont {Chen}\ \emph {et~al.}(2011)\citenamefont {Chen},
  \citenamefont {Hesse}, \citenamefont {Turschner},\ and\ \citenamefont
  {Beck}}]{chen2011improving}%
  \BibitemOpen
  \bibfield  {author} {\bibinfo {author} {\bibfnamefont {Y.}~\bibnamefont
  {Chen}}, \bibinfo {author} {\bibfnamefont {R.}~\bibnamefont {Hesse}},
  \bibinfo {author} {\bibfnamefont {D.}~\bibnamefont {Turschner}},\ and\
  \bibinfo {author} {\bibfnamefont {H.-P.}\ \bibnamefont {Beck}},\ }in\
  \href@noop {} {\emph {\bibinfo {booktitle} {2011 International Conference on
  Power Engineering, Energy and Electrical Drives}}}\ (\bibinfo {organization}
  {IEEE},\ \bibinfo {year} {2011})\ pp.\ \bibinfo {pages} {1--6}\BibitemShut
  {NoStop}%
\bibitem [{\citenamefont {Chen}\ \emph {et~al.}(2012)\citenamefont {Chen},
  \citenamefont {Hesse}, \citenamefont {Turschner},\ and\ \citenamefont
  {Beck}}]{chen2012comparison}%
  \BibitemOpen
  \bibfield  {author} {\bibinfo {author} {\bibfnamefont {Y.}~\bibnamefont
  {Chen}}, \bibinfo {author} {\bibfnamefont {R.}~\bibnamefont {Hesse}},
  \bibinfo {author} {\bibfnamefont {D.}~\bibnamefont {Turschner}},\ and\
  \bibinfo {author} {\bibfnamefont {H.-P.}\ \bibnamefont {Beck}},\ }in\
  \href@noop {} {\emph {\bibinfo {booktitle} {International conference on
  renewable energies and power quality}}},\ Vol.~\bibinfo {volume} {1}\
  (\bibinfo {year} {2012})\ pp.\ \bibinfo {pages} {414--424}\BibitemShut
  {NoStop}%
\bibitem [{\citenamefont {{Schiffer}}\ \emph {et~al.}(2014)\citenamefont
  {{Schiffer}}, \citenamefont {{Ortega}}, \citenamefont {{Astolfi}},
  \citenamefont {{Raisch}},\ and\ \citenamefont {{Sezi}}}]{schiffer2014}%
  \BibitemOpen
  \bibfield  {author} {\bibinfo {author} {\bibfnamefont {J.}~\bibnamefont
  {{Schiffer}}}, \bibinfo {author} {\bibfnamefont {R.}~\bibnamefont
  {{Ortega}}}, \bibinfo {author} {\bibfnamefont {A.}~\bibnamefont {{Astolfi}}},
  \bibinfo {author} {\bibfnamefont {J.}~\bibnamefont {{Raisch}}},\ and\
  \bibinfo {author} {\bibfnamefont {T.}~\bibnamefont {{Sezi}}},\ }\href
  {https://doi.org/10.1016/j.automatica.2014.08.009} {\bibfield  {journal}
  {\bibinfo  {journal} {Automatica}\ }\textbf {\bibinfo {volume} {50}},\
  \bibinfo {pages} {2457} (\bibinfo {year} {2014})}\BibitemShut {NoStop}%
\bibitem [{\citenamefont {{Seo}}\ \emph {et~al.}(2019)\citenamefont {{Seo}},
  \citenamefont {{Colombino}}, \citenamefont {{Subotic}}, \citenamefont
  {{Johnson}}, \citenamefont {{Groß}},\ and\ \citenamefont
  {{Dörfler}}}]{doerfler2018}%
  \BibitemOpen
  \bibfield  {author} {\bibinfo {author} {\bibfnamefont {G.}~\bibnamefont
  {{Seo}}}, \bibinfo {author} {\bibfnamefont {M.}~\bibnamefont {{Colombino}}},
  \bibinfo {author} {\bibfnamefont {I.}~\bibnamefont {{Subotic}}}, \bibinfo
  {author} {\bibfnamefont {B.}~\bibnamefont {{Johnson}}}, \bibinfo {author}
  {\bibfnamefont {D.}~\bibnamefont {{Groß}}},\ and\ \bibinfo {author}
  {\bibfnamefont {F.}~\bibnamefont {{Dörfler}}},\ }in\ \href
  {https://doi.org/10.1109/APEC.2019.8722028} {\emph {\bibinfo {booktitle}
  {2019 IEEE Applied Power Electronics Conference and Exposition (APEC)}}}\
  (\bibinfo {year} {2019})\ pp.\ \bibinfo {pages} {561--566}\BibitemShut
  {NoStop}%
\bibitem [{\citenamefont {{Milano}}\ \emph {et~al.}(2018)\citenamefont
  {{Milano}}, \citenamefont {{D{\"{o}}rfler}}, \citenamefont {{Hug}},
  \citenamefont {{Hill}},\ and\ \citenamefont
  {{Verbi{\v{c}}}}}]{doerfler2018a}%
  \BibitemOpen
  \bibfield  {author} {\bibinfo {author} {\bibfnamefont {F.}~\bibnamefont
  {{Milano}}}, \bibinfo {author} {\bibfnamefont {F.}~\bibnamefont
  {{D{\"{o}}rfler}}}, \bibinfo {author} {\bibfnamefont {G.}~\bibnamefont
  {{Hug}}}, \bibinfo {author} {\bibfnamefont {D.~J.}\ \bibnamefont {{Hill}}},\
  and\ \bibinfo {author} {\bibfnamefont {G.}~\bibnamefont {{Verbi{\v{c}}}}},\
  }in\ \href {https://doi.org/10.23919/PSCC.2018.8450880} {\emph {\bibinfo
  {booktitle} {2018 Power Systems Computation Conference (PSCC)}}}\ (\bibinfo
  {year} {2018})\ pp.\ \bibinfo {pages} {1--25}\BibitemShut {NoStop}%
\bibitem [{\citenamefont {{Gro{\ss}}}\ \emph {et~al.}(2019)\citenamefont
  {{Gro{\ss}}}, \citenamefont {{Colombino}}, \citenamefont {{Brouillon}},\ and\
  \citenamefont {{D{\"{o}}rfler}}}]{doerfler2019}%
  \BibitemOpen
  \bibfield  {author} {\bibinfo {author} {\bibfnamefont {D.}~\bibnamefont
  {{Gro{\ss}}}}, \bibinfo {author} {\bibfnamefont {M.}~\bibnamefont
  {{Colombino}}}, \bibinfo {author} {\bibfnamefont {J.}~\bibnamefont
  {{Brouillon}}},\ and\ \bibinfo {author} {\bibfnamefont {F.}~\bibnamefont
  {{D{\"{o}}rfler}}},\ }\href {https://doi.org/10.1109/TCNS.2019.2921347}
  {\bibfield  {journal} {\bibinfo  {journal} {IEEE Transactions on Control of
  Network Systems}\ }\textbf {\bibinfo {volume} {6}},\ \bibinfo {pages} {1148}
  (\bibinfo {year} {2019})}\BibitemShut {NoStop}%
\bibitem [{\citenamefont {{Haehne}}\ \emph {et~al.}(2018)\citenamefont
  {{Haehne}}, \citenamefont {{Schottler}}, \citenamefont {{Waechter}},
  \citenamefont {{Peinke}},\ and\ \citenamefont {{Kamps}}}]{haehne2018}%
  \BibitemOpen
  \bibfield  {author} {\bibinfo {author} {\bibfnamefont {H.}~\bibnamefont
  {{Haehne}}}, \bibinfo {author} {\bibfnamefont {J.}~\bibnamefont
  {{Schottler}}}, \bibinfo {author} {\bibfnamefont {M.}~\bibnamefont
  {{Waechter}}}, \bibinfo {author} {\bibfnamefont {J.}~\bibnamefont
  {{Peinke}}},\ and\ \bibinfo {author} {\bibfnamefont {O.}~\bibnamefont
  {{Kamps}}},\ }\href {https://doi.org/10.1209/0295-5075/121/30001} {\bibfield
  {journal} {\bibinfo  {journal} {{EPL} (Europhysics Letters)}\ }\textbf
  {\bibinfo {volume} {121}},\ \bibinfo {pages} {30001} (\bibinfo {year}
  {2018})}\BibitemShut {NoStop}%
\bibitem [{\citenamefont {{Haehne}}\ \emph {et~al.}(2019)\citenamefont
  {{Haehne}}, \citenamefont {{Schmietendorf}}, \citenamefont {{Tamrakar}},
  \citenamefont {{Peinke}},\ and\ \citenamefont {{Kettemann}}}]{haehne2019}%
  \BibitemOpen
  \bibfield  {author} {\bibinfo {author} {\bibfnamefont {H.}~\bibnamefont
  {{Haehne}}}, \bibinfo {author} {\bibfnamefont {K.}~\bibnamefont
  {{Schmietendorf}}}, \bibinfo {author} {\bibfnamefont {S.}~\bibnamefont
  {{Tamrakar}}}, \bibinfo {author} {\bibfnamefont {J.}~\bibnamefont
  {{Peinke}}},\ and\ \bibinfo {author} {\bibfnamefont {S.}~\bibnamefont
  {{Kettemann}}},\ }\href {https://doi.org/10.1103/PhysRevE.99.050301}
  {\bibfield  {journal} {\bibinfo  {journal} {Phys. Rev. E}\ }\textbf {\bibinfo
  {volume} {99}},\ \bibinfo {pages} {050301} (\bibinfo {year}
  {2019})}\BibitemShut {NoStop}%
\bibitem [{\citenamefont {{Muratori}}(2018)}]{muratori2018}%
  \BibitemOpen
  \bibfield  {author} {\bibinfo {author} {\bibfnamefont {M.}~\bibnamefont
  {{Muratori}}},\ }\bibfield  {journal} {\bibinfo  {journal} {Nature Energy}\
  }\textbf {\bibinfo {volume} {3}},\ \href
  {https://doi.org/10.1038/s41560-017-0074-z} {10.1038/s41560-017-0074-z}
  (\bibinfo {year} {2018})\BibitemShut {NoStop}%
\bibitem [{\citenamefont {{Anvari}}\ \emph {et~al.}(2020)\citenamefont
  {{Anvari}}, \citenamefont {{Proedrou}}, \citenamefont {{Schaefer}},
  \citenamefont {{Beck}}, \citenamefont {{Kantz}},\ and\ \citenamefont
  {{Timme}}}]{anvari2020}%
  \BibitemOpen
  \bibfield  {author} {\bibinfo {author} {\bibfnamefont {M.}~\bibnamefont
  {{Anvari}}}, \bibinfo {author} {\bibfnamefont {E.}~\bibnamefont
  {{Proedrou}}}, \bibinfo {author} {\bibfnamefont {B.}~\bibnamefont
  {{Schaefer}}}, \bibinfo {author} {\bibfnamefont {C.}~\bibnamefont {{Beck}}},
  \bibinfo {author} {\bibfnamefont {H.}~\bibnamefont {{Kantz}}},\ and\ \bibinfo
  {author} {\bibfnamefont {M.}~\bibnamefont {{Timme}}},\ }\href@noop {}
  {\bibinfo {title} {Data-driven load profiles and the dynamics of residential
  electric power consumption}} (\bibinfo {year} {2020}),\ \Eprint
  {https://arxiv.org/abs/2009.09287} {arXiv:2009.09287 [physics.app-ph]}
  \BibitemShut {NoStop}%
\bibitem [{\citenamefont {{Kuznetsov}}(1998)}]{kuznetsov1998}%
  \BibitemOpen
  \bibfield  {author} {\bibinfo {author} {\bibfnamefont {Y.~A.}\ \bibnamefont
  {{Kuznetsov}}},\ }\href@noop {} {\emph {\bibinfo {title} {Elements of Applied
  Bifurcation Theory (2\textsuperscript{nd} Ed.)}}}\ (\bibinfo  {publisher}
  {Springer-Verlag},\ \bibinfo {address} {Berlin, Heidelberg},\ \bibinfo {year}
  {1998})\BibitemShut {NoStop}%
\bibitem [{\citenamefont {{Schiffer}}\ \emph {et~al.}(2016)\citenamefont
  {{Schiffer}}, \citenamefont {{Zonetti}}, \citenamefont {{Ortega}},
  \citenamefont {{Stankovi{\'{c}}}}, \citenamefont {{Sezi}},\ and\
  \citenamefont {{Raisch}}}]{schiffer2016}%
  \BibitemOpen
  \bibfield  {author} {\bibinfo {author} {\bibfnamefont {J.}~\bibnamefont
  {{Schiffer}}}, \bibinfo {author} {\bibfnamefont {D.}~\bibnamefont
  {{Zonetti}}}, \bibinfo {author} {\bibfnamefont {R.}~\bibnamefont {{Ortega}}},
  \bibinfo {author} {\bibfnamefont {A.~M.}\ \bibnamefont {{Stankovi{\'{c}}}}},
  \bibinfo {author} {\bibfnamefont {T.}~\bibnamefont {{Sezi}}},\ and\ \bibinfo
  {author} {\bibfnamefont {J.}~\bibnamefont {{Raisch}}},\ }\href
  {https://doi.org/10.1016/j.automatica.2016.07.036} {\bibfield  {journal}
  {\bibinfo  {journal} {Automatica}\ }\textbf {\bibinfo {volume} {74}},\
  \bibinfo {pages} {135} (\bibinfo {year} {2016})}\BibitemShut {NoStop}%
\bibitem [{\citenamefont {{Curi}}\ \emph {et~al.}(2017)\citenamefont {{Curi}},
  \citenamefont {{Gro{\ss}}},\ and\ \citenamefont
  {{D{\"{o}}rfler}}}]{doerfler2017}%
  \BibitemOpen
  \bibfield  {author} {\bibinfo {author} {\bibfnamefont {S.}~\bibnamefont
  {{Curi}}}, \bibinfo {author} {\bibfnamefont {D.}~\bibnamefont {{Gro{\ss}}}},\
  and\ \bibinfo {author} {\bibfnamefont {F.}~\bibnamefont {{D{\"{o}}rfler}}},\
  }in\ \href@noop {} {\emph {\bibinfo {booktitle} {2017 IEEE
  56\textsuperscript{th} Annual Conference on Decision and Control (CDC)}}}\
  (\bibinfo {year} {2017})\ pp.\ \bibinfo {pages} {5708--5713}\BibitemShut
  {NoStop}%
\bibitem [{\citenamefont {Duesterhoeft}\ \emph {et~al.}(1951)\citenamefont
  {Duesterhoeft}, \citenamefont {Schulz},\ and\ \citenamefont
  {Clarke}}]{duesterhoeft1951determination}%
  \BibitemOpen
  \bibfield  {author} {\bibinfo {author} {\bibfnamefont {W.}~\bibnamefont
  {Duesterhoeft}}, \bibinfo {author} {\bibfnamefont {M.~W.}\ \bibnamefont
  {Schulz}},\ and\ \bibinfo {author} {\bibfnamefont {E.}~\bibnamefont
  {Clarke}},\ }\href@noop {} {\bibfield  {journal} {\bibinfo  {journal}
  {Transactions of the American Institute of Electrical Engineers}\ }\textbf
  {\bibinfo {volume} {70}},\ \bibinfo {pages} {1248} (\bibinfo {year}
  {1951})}\BibitemShut {NoStop}%
\bibitem [{\citenamefont {Remmert}(1991)}]{remmert1991theory}%
  \BibitemOpen
  \bibfield  {author} {\bibinfo {author} {\bibfnamefont {R.}~\bibnamefont
  {Remmert}},\ }\href@noop {} {\emph {\bibinfo {title} {Theory of complex
  functions}}},\ Vol.\ \bibinfo {volume} {122}\ (\bibinfo  {publisher}
  {Springer Science \& Business Media},\ \bibinfo {year} {1991})\BibitemShut
  {NoStop}%
\bibitem [{\citenamefont {{Erickson}}\ and\ \citenamefont
  {{Maksimovic}}(2001)}]{erickson2001}%
  \BibitemOpen
  \bibfield  {author} {\bibinfo {author} {\bibfnamefont {R.~W.}\ \bibnamefont
  {{Erickson}}}\ and\ \bibinfo {author} {\bibfnamefont {D.}~\bibnamefont
  {{Maksimovic}}},\ }\href {https://doi.org/10.1007/b100747} {\emph {\bibinfo
  {title} {Fundamentals of Power Electronics}}},\ \bibinfo {edition} {2nd}\
  ed.\ (\bibinfo  {publisher} {Springer US},\ \bibinfo {year}
  {2001})\BibitemShut {NoStop}%
\bibitem [{\citenamefont {{Chiniforoosh}}\ \emph {et~al.}(2010)\citenamefont
  {{Chiniforoosh}}, \citenamefont {{Jatskevich}}, \citenamefont {{Yazdani}},
  \citenamefont {{Sood}}, \citenamefont {{Dinavahi}}, \citenamefont
  {{Martinez}},\ and\ \citenamefont {{Ramirez}}}]{chiniforoosh2010}%
  \BibitemOpen
  \bibfield  {author} {\bibinfo {author} {\bibfnamefont {S.}~\bibnamefont
  {{Chiniforoosh}}}, \bibinfo {author} {\bibfnamefont {J.}~\bibnamefont
  {{Jatskevich}}}, \bibinfo {author} {\bibfnamefont {A.}~\bibnamefont
  {{Yazdani}}}, \bibinfo {author} {\bibfnamefont {V.}~\bibnamefont {{Sood}}},
  \bibinfo {author} {\bibfnamefont {V.}~\bibnamefont {{Dinavahi}}}, \bibinfo
  {author} {\bibfnamefont {J.~A.}\ \bibnamefont {{Martinez}}},\ and\ \bibinfo
  {author} {\bibfnamefont {A.}~\bibnamefont {{Ramirez}}},\ }\href
  {https://doi.org/10.1109/TPWRD.2010.2043859} {\bibfield  {journal} {\bibinfo
  {journal} {IEEE Transactions on Power Delivery}\ }\textbf {\bibinfo {volume}
  {25}},\ \bibinfo {pages} {2655} (\bibinfo {year} {2010})}\BibitemShut
  {NoStop}%
\bibitem [{\citenamefont {{Gro{\ss}}}\ and\ \citenamefont
  {{D{\"{o}}rfler}}(2019)}]{doerfler2019b}%
  \BibitemOpen
  \bibfield  {author} {\bibinfo {author} {\bibfnamefont {D.}~\bibnamefont
  {{Gro{\ss}}}}\ and\ \bibinfo {author} {\bibfnamefont {F.}~\bibnamefont
  {{D{\"{o}}rfler}}},\ }in\ \href@noop {} {\emph {\bibinfo {booktitle} {2019
  57\textsuperscript{th} Annual Allerton Conference on Communication, Control,
  and Computing (Allerton)}}}\ (\bibinfo {year} {2019})\ pp.\ \bibinfo {pages}
  {326--333}\BibitemShut {NoStop}%
\bibitem [{\citenamefont {Coulson}\ \emph {et~al.}(2019)\citenamefont
  {Coulson}, \citenamefont {Lygeros},\ and\ \citenamefont
  {D{\"o}rfler}}]{coulson2019data}%
  \BibitemOpen
  \bibfield  {author} {\bibinfo {author} {\bibfnamefont {J.}~\bibnamefont
  {Coulson}}, \bibinfo {author} {\bibfnamefont {J.}~\bibnamefont {Lygeros}},\
  and\ \bibinfo {author} {\bibfnamefont {F.}~\bibnamefont {D{\"o}rfler}},\ }in\
  \href@noop {} {\emph {\bibinfo {booktitle} {2019 18th European Control
  Conference (ECC)}}}\ (\bibinfo {organization} {IEEE},\ \bibinfo {year}
  {2019})\ pp.\ \bibinfo {pages} {307--312}\BibitemShut {NoStop}%
\bibitem [{\citenamefont {{Olver}}(1993)}]{olver1993}%
  \BibitemOpen
  \bibfield  {author} {\bibinfo {author} {\bibfnamefont {P.}~\bibnamefont
  {{Olver}}},\ }\href@noop {} {\emph {\bibinfo {title} {Applications of Lie
  Groups to Differential Equations}}},\ \bibinfo {edition} {2nd}\ ed.\
  (\bibinfo  {publisher} {Springer New York},\ \bibinfo {year}
  {1993})\BibitemShut {NoStop}%
\bibitem [{\citenamefont {{Johnson}}\ \emph {et~al.}(2016)\citenamefont
  {{Johnson}}, \citenamefont {{Sinha}}, \citenamefont {{Ainsworth}},
  \citenamefont {{Dörfler}},\ and\ \citenamefont {{Dhople}}}]{johnson2016}%
  \BibitemOpen
  \bibfield  {author} {\bibinfo {author} {\bibfnamefont {B.~B.}\ \bibnamefont
  {{Johnson}}}, \bibinfo {author} {\bibfnamefont {M.}~\bibnamefont {{Sinha}}},
  \bibinfo {author} {\bibfnamefont {N.~G.}\ \bibnamefont {{Ainsworth}}},
  \bibinfo {author} {\bibfnamefont {F.}~\bibnamefont {{Dörfler}}},\ and\
  \bibinfo {author} {\bibfnamefont {S.~V.}\ \bibnamefont {{Dhople}}},\ }\href
  {https://doi.org/10.1109/TPEL.2015.2497217} {\bibfield  {journal} {\bibinfo
  {journal} {IEEE Transactions on Power Electronics}\ }\textbf {\bibinfo
  {volume} {31}},\ \bibinfo {pages} {6002} (\bibinfo {year}
  {2016})}\BibitemShut {NoStop}%
\bibitem [{\citenamefont {{Johnson}}\ \emph {et~al.}(2017)\citenamefont
  {{Johnson}}, \citenamefont {{Rodriguez}}, \citenamefont {{Sinha}},\ and\
  \citenamefont {{Dhople}}}]{johnson2017}%
  \BibitemOpen
  \bibfield  {author} {\bibinfo {author} {\bibfnamefont {B.}~\bibnamefont
  {{Johnson}}}, \bibinfo {author} {\bibfnamefont {M.}~\bibnamefont
  {{Rodriguez}}}, \bibinfo {author} {\bibfnamefont {M.}~\bibnamefont
  {{Sinha}}},\ and\ \bibinfo {author} {\bibfnamefont {S.}~\bibnamefont
  {{Dhople}}},\ }in\ \href {https://doi.org/10.1109/COMPEL.2017.8013298} {\emph
  {\bibinfo {booktitle} {2017 IEEE 18th Workshop on Control and Modeling for
  Power Electronics (COMPEL)}}}\ (\bibinfo {year} {2017})\ pp.\ \bibinfo
  {pages} {1--6}\BibitemShut {NoStop}%
\bibitem [{\citenamefont {{Rodrigues}}\ \emph {et~al.}(2016)\citenamefont
  {{Rodrigues}}, \citenamefont {{Peron}}, \citenamefont {{Ji}},\ and\
  \citenamefont {{Kurths}}}]{rodrigues2016}%
  \BibitemOpen
  \bibfield  {author} {\bibinfo {author} {\bibfnamefont {F.}~\bibnamefont
  {{Rodrigues}}}, \bibinfo {author} {\bibfnamefont {T.}~\bibnamefont
  {{Peron}}}, \bibinfo {author} {\bibfnamefont {P.}~\bibnamefont {{Ji}}},\ and\
  \bibinfo {author} {\bibfnamefont {J.}~\bibnamefont {{Kurths}}},\ }\href
  {https://doi.org/10.1016/j.physrep.2015.10.008} {\bibfield  {journal}
  {\bibinfo  {journal} {Physics Reports}\ }\textbf {\bibinfo {volume} {610}},\
  \bibinfo {pages} {1} (\bibinfo {year} {2016})}\BibitemShut {NoStop}%
\bibitem [{\citenamefont {Simpson-Porco}\ \emph {et~al.}(2013)\citenamefont
  {Simpson-Porco}, \citenamefont {D{\"o}rfler},\ and\ \citenamefont
  {Bullo}}]{simpson2013synchronization}%
  \BibitemOpen
  \bibfield  {author} {\bibinfo {author} {\bibfnamefont {J.~W.}\ \bibnamefont
  {Simpson-Porco}}, \bibinfo {author} {\bibfnamefont {F.}~\bibnamefont
  {D{\"o}rfler}},\ and\ \bibinfo {author} {\bibfnamefont {F.}~\bibnamefont
  {Bullo}},\ }\href@noop {} {\bibfield  {journal} {\bibinfo  {journal}
  {Automatica}\ }\textbf {\bibinfo {volume} {49}},\ \bibinfo {pages} {2603}
  (\bibinfo {year} {2013})}\BibitemShut {NoStop}%
\bibitem [{\citenamefont {{Machowski}and Z.~{Lubo{\'s}ny}}\ \emph
  {et~al.}(2020)\citenamefont {{Machowski}and Z.~{Lubo{\'s}ny}}, \citenamefont
  {{Bia{\l}ek}},\ and\ \citenamefont {{Bumby}}}]{machowski2020}%
  \BibitemOpen
  \bibfield  {author} {\bibinfo {author} {\bibfnamefont {J.}~\bibnamefont
  {{Machowski}and Z.~{Lubo{\'s}ny}}}, \bibinfo {author} {\bibfnamefont
  {J.}~\bibnamefont {{Bia{\l}ek}}},\ and\ \bibinfo {author} {\bibfnamefont
  {J.}~\bibnamefont {{Bumby}}},\ }\href@noop {} {\emph {\bibinfo {title} {Power
  System Dynamics. Stability and Control.}}},\ \bibinfo {edition} {3rd}\ ed.\
  (\bibinfo  {publisher} {WILEY},\ \bibinfo {year} {2020})\BibitemShut
  {NoStop}%
\bibitem [{\citenamefont {{Monshizadeh}}\ \emph {et~al.}(2016)\citenamefont
  {{Monshizadeh}}, \citenamefont {{De Persis}}, \citenamefont {{Monshizadeh}},\
  and\ \citenamefont {{van der Schaft}}}]{monshizadeh2016}%
  \BibitemOpen
  \bibfield  {author} {\bibinfo {author} {\bibfnamefont {P.}~\bibnamefont
  {{Monshizadeh}}}, \bibinfo {author} {\bibfnamefont {C.}~\bibnamefont {{De
  Persis}}}, \bibinfo {author} {\bibfnamefont {N.}~\bibnamefont
  {{Monshizadeh}}},\ and\ \bibinfo {author} {\bibfnamefont {A.~J.}\
  \bibnamefont {{van der Schaft}}},\ }in\ \href@noop {} {\emph {\bibinfo
  {booktitle} {2016 IEEE 55\textsuperscript{th} Conference on Decision and
  Control (CDC)}}}\ (\bibinfo {year} {2016})\ pp.\ \bibinfo {pages}
  {4116--4121}\BibitemShut {NoStop}%
\bibitem [{\citenamefont {{Schmietendorf}}\ \emph {et~al.}(2013)\citenamefont
  {{Schmietendorf}}, \citenamefont {{Peinke}}, \citenamefont {{Friedrich}},\
  and\ \citenamefont {{Kamps}}}]{schmietendorf2013}%
  \BibitemOpen
  \bibfield  {author} {\bibinfo {author} {\bibfnamefont {K.}~\bibnamefont
  {{Schmietendorf}}}, \bibinfo {author} {\bibfnamefont {J.}~\bibnamefont
  {{Peinke}}}, \bibinfo {author} {\bibfnamefont {R.}~\bibnamefont
  {{Friedrich}}},\ and\ \bibinfo {author} {\bibfnamefont {O.}~\bibnamefont
  {{Kamps}}},\ }\bibfield  {journal} {\bibinfo  {journal} {The European
  Physical Journal Special Topics}\ }\textbf {\bibinfo {volume} {223}},\ \href
  {https://doi.org/10.1140/epjst/e2014-02209-8} {10.1140/epjst/e2014-02209-8}
  (\bibinfo {year} {2013})\BibitemShut {NoStop}%
\bibitem [{\citenamefont {{Kuramoto}}(1984)}]{kuramoto1984}%
  \BibitemOpen
  \bibfield  {author} {\bibinfo {author} {\bibfnamefont {Y.}~\bibnamefont
  {{Kuramoto}}},\ }\href@noop {} {\emph {\bibinfo {title} {Chemical
  Oscillations, Waves, and Turbulence}}}\ (\bibinfo  {publisher}
  {Springer-Verlag Berlin Heidelberg},\ \bibinfo {year} {1984})\BibitemShut
  {NoStop}%
\bibitem [{\citenamefont {{Panteley}}\ \emph {et~al.}(2015)\citenamefont
  {{Panteley}}, \citenamefont {{Loria}},\ and\ \citenamefont
  {{Ati}}}]{panteley2015}%
  \BibitemOpen
  \bibfield  {author} {\bibinfo {author} {\bibfnamefont {E.}~\bibnamefont
  {{Panteley}}}, \bibinfo {author} {\bibfnamefont {A.}~\bibnamefont
  {{Loria}}},\ and\ \bibinfo {author} {\bibfnamefont {A.~E.}\ \bibnamefont
  {{Ati}}},\ }\href {https://doi.org/10.1016/j.ifacol.2015.09.260} {\bibfield
  {journal} {\bibinfo  {journal} {IFAC-PapersOnLine}\ }\textbf {\bibinfo
  {volume} {48}},\ \bibinfo {pages} {645 } (\bibinfo {year}
  {2015})}\BibitemShut {NoStop}%
\bibitem [{\citenamefont {{Maghenem}}\ \emph {et~al.}(2016)\citenamefont
  {{Maghenem}}, \citenamefont {{Panteley}},\ and\ \citenamefont
  {{Loría}}}]{panteley2016}%
  \BibitemOpen
  \bibfield  {author} {\bibinfo {author} {\bibfnamefont {M.}~\bibnamefont
  {{Maghenem}}}, \bibinfo {author} {\bibfnamefont {E.}~\bibnamefont
  {{Panteley}}},\ and\ \bibinfo {author} {\bibfnamefont {A.}~\bibnamefont
  {{Loría}}},\ }in\ \href {https://doi.org/10.1109/CDC.2016.7798651} {\emph
  {\bibinfo {booktitle} {2016 IEEE 55th Conference on Decision and Control
  (CDC)}}}\ (\bibinfo {year} {2016})\ pp.\ \bibinfo {pages}
  {2581--2586}\BibitemShut {NoStop}%
\bibitem [{\citenamefont {{Röhm}}\ \emph {et~al.}(2018)\citenamefont
  {{Röhm}}, \citenamefont {{Lüdge}},\ and\ \citenamefont
  {{Schneider}}}]{roehm2018}%
  \BibitemOpen
  \bibfield  {author} {\bibinfo {author} {\bibfnamefont {A.}~\bibnamefont
  {{Röhm}}}, \bibinfo {author} {\bibfnamefont {K.}~\bibnamefont {{Lüdge}}},\
  and\ \bibinfo {author} {\bibfnamefont {I.}~\bibnamefont {{Schneider}}},\
  }\href {https://doi.org/10.1063/1.5018262} {\bibfield  {journal} {\bibinfo
  {journal} {Chaos (Woodbury, N.Y.)}\ }\textbf {\bibinfo {volume} {28}},\
  \bibinfo {pages} {063114} (\bibinfo {year} {2018})}\BibitemShut {NoStop}%
\bibitem [{\citenamefont {Planas}\ \emph {et~al.}(2013)\citenamefont {Planas},
  \citenamefont {Gil-de Muro}, \citenamefont {Andreu}, \citenamefont
  {Kortabarria},\ and\ \citenamefont {de~Alegr{\'\i}a}}]{planas2013design}%
  \BibitemOpen
  \bibfield  {author} {\bibinfo {author} {\bibfnamefont {E.}~\bibnamefont
  {Planas}}, \bibinfo {author} {\bibfnamefont {A.}~\bibnamefont {Gil-de Muro}},
  \bibinfo {author} {\bibfnamefont {J.}~\bibnamefont {Andreu}}, \bibinfo
  {author} {\bibfnamefont {I.}~\bibnamefont {Kortabarria}},\ and\ \bibinfo
  {author} {\bibfnamefont {I.~M.}\ \bibnamefont {de~Alegr{\'\i}a}},\
  }\href@noop {} {\bibfield  {journal} {\bibinfo  {journal} {IET Renewable
  Power Generation}\ }\textbf {\bibinfo {volume} {7}},\ \bibinfo {pages} {458}
  (\bibinfo {year} {2013})}\BibitemShut {NoStop}%
\bibitem [{\citenamefont {Plietzsch}\ \emph {et~al.}(2021)\citenamefont
  {Plietzsch}, \citenamefont {Kogler}, \citenamefont {Auer}, \citenamefont
  {Merino}, \citenamefont {Gil-de Muro}, \citenamefont {Li{\ss}e},
  \citenamefont {Vogel},\ and\ \citenamefont
  {Hellmann}}]{plietzsch2021powerdynamics}%
  \BibitemOpen
  \bibfield  {author} {\bibinfo {author} {\bibfnamefont {A.}~\bibnamefont
  {Plietzsch}}, \bibinfo {author} {\bibfnamefont {R.}~\bibnamefont {Kogler}},
  \bibinfo {author} {\bibfnamefont {S.}~\bibnamefont {Auer}}, \bibinfo {author}
  {\bibfnamefont {J.}~\bibnamefont {Merino}}, \bibinfo {author} {\bibfnamefont
  {A.}~\bibnamefont {Gil-de Muro}}, \bibinfo {author} {\bibfnamefont
  {J.}~\bibnamefont {Li{\ss}e}}, \bibinfo {author} {\bibfnamefont
  {C.}~\bibnamefont {Vogel}},\ and\ \bibinfo {author} {\bibfnamefont
  {F.}~\bibnamefont {Hellmann}},\ }\href@noop {} {\bibfield  {journal}
  {\bibinfo  {journal} {arXiv preprint arXiv:2101.02103}\ } (\bibinfo {year}
  {2021})}\BibitemShut {NoStop}%
\bibitem [{\citenamefont {Canizares}\ and\ \citenamefont
  {Kodsi}(2003)}]{canizares2003modeling}%
  \BibitemOpen
  \bibfield  {author} {\bibinfo {author} {\bibfnamefont {C.}~\bibnamefont
  {Canizares}}\ and\ \bibinfo {author} {\bibfnamefont {S.}~\bibnamefont
  {Kodsi}},\ }\href@noop {} {\bibfield  {journal} {\bibinfo  {journal}
  {Technical Report 2003-3}\ } (\bibinfo {year} {2003})}\BibitemShut {NoStop}%
\bibitem [{\citenamefont {Menck}\ \emph {et~al.}(2013)\citenamefont {Menck},
  \citenamefont {Heitzig}, \citenamefont {Marwan},\ and\ \citenamefont
  {Kurths}}]{menck2013basin}%
  \BibitemOpen
  \bibfield  {author} {\bibinfo {author} {\bibfnamefont {P.~J.}\ \bibnamefont
  {Menck}}, \bibinfo {author} {\bibfnamefont {J.}~\bibnamefont {Heitzig}},
  \bibinfo {author} {\bibfnamefont {N.}~\bibnamefont {Marwan}},\ and\ \bibinfo
  {author} {\bibfnamefont {J.}~\bibnamefont {Kurths}},\ }\href@noop {}
  {\bibfield  {journal} {\bibinfo  {journal} {Nature physics}\ }\textbf
  {\bibinfo {volume} {9}},\ \bibinfo {pages} {89} (\bibinfo {year}
  {2013})}\BibitemShut {NoStop}%
\bibitem [{\citenamefont {Menck}\ \emph {et~al.}(2014)\citenamefont {Menck},
  \citenamefont {Heitzig}, \citenamefont {Kurths},\ and\ \citenamefont
  {Schellnhuber}}]{menck2014dead}%
  \BibitemOpen
  \bibfield  {author} {\bibinfo {author} {\bibfnamefont {P.~J.}\ \bibnamefont
  {Menck}}, \bibinfo {author} {\bibfnamefont {J.}~\bibnamefont {Heitzig}},
  \bibinfo {author} {\bibfnamefont {J.}~\bibnamefont {Kurths}},\ and\ \bibinfo
  {author} {\bibfnamefont {H.~J.}\ \bibnamefont {Schellnhuber}},\ }\href@noop
  {} {\bibfield  {journal} {\bibinfo  {journal} {Nature communications}\
  }\textbf {\bibinfo {volume} {5}},\ \bibinfo {pages} {1} (\bibinfo {year}
  {2014})}\BibitemShut {NoStop}%
\bibitem [{\citenamefont {Brummitt}\ \emph {et~al.}(2013)\citenamefont
  {Brummitt}, \citenamefont {Hines}, \citenamefont {Dobson}, \citenamefont
  {Moore},\ and\ \citenamefont {D'Souza}}]{brummitt2013transdisciplinary}%
  \BibitemOpen
  \bibfield  {author} {\bibinfo {author} {\bibfnamefont {C.~D.}\ \bibnamefont
  {Brummitt}}, \bibinfo {author} {\bibfnamefont {P.~D.}\ \bibnamefont {Hines}},
  \bibinfo {author} {\bibfnamefont {I.}~\bibnamefont {Dobson}}, \bibinfo
  {author} {\bibfnamefont {C.}~\bibnamefont {Moore}},\ and\ \bibinfo {author}
  {\bibfnamefont {R.~M.}\ \bibnamefont {D'Souza}},\ }\href@noop {} {\bibfield
  {journal} {\bibinfo  {journal} {Proceedings of the National Academy of
  Sciences}\ }\textbf {\bibinfo {volume} {110}},\ \bibinfo {pages} {12159}
  (\bibinfo {year} {2013})}\BibitemShut {NoStop}%
\bibitem [{\citenamefont {Auer}\ \emph {et~al.}(2016)\citenamefont {Auer},
  \citenamefont {Kleis}, \citenamefont {Schultz}, \citenamefont {Kurths},\ and\
  \citenamefont {Hellmann}}]{auer2016impact}%
  \BibitemOpen
  \bibfield  {author} {\bibinfo {author} {\bibfnamefont {S.}~\bibnamefont
  {Auer}}, \bibinfo {author} {\bibfnamefont {K.}~\bibnamefont {Kleis}},
  \bibinfo {author} {\bibfnamefont {P.}~\bibnamefont {Schultz}}, \bibinfo
  {author} {\bibfnamefont {J.}~\bibnamefont {Kurths}},\ and\ \bibinfo {author}
  {\bibfnamefont {F.}~\bibnamefont {Hellmann}},\ }\href@noop {} {\bibfield
  {journal} {\bibinfo  {journal} {The European Physical Journal Special
  Topics}\ }\textbf {\bibinfo {volume} {225}},\ \bibinfo {pages} {609}
  (\bibinfo {year} {2016})}\BibitemShut {NoStop}%
\bibitem [{\citenamefont {Vogel}\ \emph {et~al.}(2020)\citenamefont {Vogel},
  \citenamefont {Auer}, \citenamefont {De{\ss}}, \citenamefont {Plietzsch},\
  and\ \citenamefont {Kogler}}]{VALERIA}%
  \BibitemOpen
  \bibfield  {author} {\bibinfo {author} {\bibfnamefont {C.}~\bibnamefont
  {Vogel}}, \bibinfo {author} {\bibfnamefont {S.}~\bibnamefont {Auer}},
  \bibinfo {author} {\bibfnamefont {T.}~\bibnamefont {De{\ss}}}, \bibinfo
  {author} {\bibfnamefont {A.}~\bibnamefont {Plietzsch}},\ and\ \bibinfo
  {author} {\bibfnamefont {R.}~\bibnamefont {Kogler}},\ }\href@noop {} {\emph
  {\bibinfo {title} {Validation of low-voltage energy and renewables
  integration analysis ({VALERIA})}}},\ \bibinfo {type} {Tech. Rep.}\ (\bibinfo
   {institution} {European Research Infrastructure supporting Smart Grid
  (ERIGrid)},\ \bibinfo {year} {2020})\ \bibinfo {note}
  {\url{https://erigrid.eu/wp-content/uploads/2020/06/ERIGrid_TA_VALERIA_Technical-Report_v01.pdf}}\BibitemShut
  {NoStop}%
\bibitem [{\citenamefont {Rackauckas}\ and\ \citenamefont
  {Nie}(2017)}]{rackauckas2017differentialequations}%
  \BibitemOpen
  \bibfield  {author} {\bibinfo {author} {\bibfnamefont {C.}~\bibnamefont
  {Rackauckas}}\ and\ \bibinfo {author} {\bibfnamefont {Q.}~\bibnamefont
  {Nie}},\ }\href@noop {} {\bibfield  {journal} {\bibinfo  {journal} {Journal
  of Open Research Software}\ }\textbf {\bibinfo {volume} {5}} (\bibinfo {year}
  {2017})}\BibitemShut {NoStop}%
\bibitem [{\citenamefont {Rackauckas}\ \emph {et~al.}(2019)\citenamefont
  {Rackauckas}, \citenamefont {Innes}, \citenamefont {Ma}, \citenamefont
  {Bettencourt}, \citenamefont {White},\ and\ \citenamefont
  {Dixit}}]{rackauckas2019diffeqflux}%
  \BibitemOpen
  \bibfield  {author} {\bibinfo {author} {\bibfnamefont {C.}~\bibnamefont
  {Rackauckas}}, \bibinfo {author} {\bibfnamefont {M.}~\bibnamefont {Innes}},
  \bibinfo {author} {\bibfnamefont {Y.}~\bibnamefont {Ma}}, \bibinfo {author}
  {\bibfnamefont {J.}~\bibnamefont {Bettencourt}}, \bibinfo {author}
  {\bibfnamefont {L.}~\bibnamefont {White}},\ and\ \bibinfo {author}
  {\bibfnamefont {V.}~\bibnamefont {Dixit}},\ }\href@noop {} {\bibfield
  {journal} {\bibinfo  {journal} {arXiv preprint arXiv:1902.02376}\ } (\bibinfo
  {year} {2019})}\BibitemShut {NoStop}%
\bibitem [{\citenamefont {{Hahn}}(1967)}]{hahn1967}%
  \BibitemOpen
  \bibfield  {author} {\bibinfo {author} {\bibfnamefont {W.}~\bibnamefont
  {{Hahn}}},\ }\href@noop {} {\emph {\bibinfo {title} {Stability of Motion}}},\
  Grundlehren der mathematischen Wissenschaften\ (\bibinfo  {publisher}
  {Springer Berlin Heidelberg},\ \bibinfo {year} {1967})\BibitemShut {NoStop}%
\bibitem [{\citenamefont {Wanner}\ and\ \citenamefont
  {Hairer}(1996)}]{wanner1996solving}%
  \BibitemOpen
  \bibfield  {author} {\bibinfo {author} {\bibfnamefont {G.}~\bibnamefont
  {Wanner}}\ and\ \bibinfo {author} {\bibfnamefont {E.}~\bibnamefont
  {Hairer}},\ }\href@noop {} {\emph {\bibinfo {title} {Solving ordinary
  differential equations II}}},\ Vol.\ \bibinfo {volume} {375}\ (\bibinfo
  {publisher} {Springer Berlin Heidelberg},\ \bibinfo {year}
  {1996})\BibitemShut {NoStop}%
\end{thebibliography}%

\newpage 
\appendix

\section{Line dynamics}\label{sec:appendix_line_dynamics}

As an addendum to section~\ref{sec:modeling} we want to briefly discuss the applicable line models for our normal form. As we take nodes to be a voltage that reacts to a current, that is, it behaves like a capacitor, our lines need to be modeled as providing a current in reaction to the terminal voltages, thus behaving like inductances.

Using the same notation as in the main section, we only need to include further dynamical variables corresponding to the currents flowing on each transmission line denoted by $j_{e,m}(t): \mathbb{R} \to \mathbb{C}$. From the standard laws of electrical circuit elements a line modeled as a resistor and an inductance in series has the equation
\begin{equation}\label{eq:current_dynamics}\begin{aligned}
	\ell_m\frac{\mathrm{d}}{\mathrm{d}t} j_{e,m} &= -r_m j_{e,m} + \sum_{n=1}^N B_{nm} u_n \;,\\
	j_n &= \sum_{m=1}^M B_{nm} j_{e,m} \;,
\end{aligned}\end{equation}
which would replace the algebraic relationship between voltage and current, eqn.~\eqref{eq:ss_algebraic}, that we used in the main section. The latter is actually derived by considering the quasi-steady state where $j_{e,m}(t) \sim \exp{i\Omega_s t}$, i.e. setting
\begin{equation*}\label{eq:steady_state}\begin{aligned}
	\frac{\mathrm{d}}{\mathrm{d}t} j_{e,m}^s &= i\Omega^s j_{e,m}^s.
\end{aligned}\end{equation*}

More sophisticated models of lines, which include line capacitances, can be naturally coupled to our nodal ODEs as long as they provide ODEs for the terminal currents. This is the case, for example for $\tau$-models and iterated $\tau$-models of transmission lines. Models such as the $\pi$-model which have ODEs for the terminal voltages lead to algebraic constraints on the system.

We also want to note that in the special case of a uniform ratio between resistance and inductance across the whole network, i.e. $\ell_m/r_m =: \tau ~\forall m \in \mathcal{E}$, the current dynamics may be directly expressed in terms of the admittance matrix $Y$ and also greatly simplified with respect to dimensionality by eliminating the line currents and writing the nodal current dynamics as
\begin{equation*}
	\frac{\mathrm{d}}{\mathrm{d}t} j_n = -\frac{1}{\tau} j_n + \left( \frac{1}{\tau} + i\Omega^s \right) \sum_{m=1}^N Y_{nm} u_m \;.
\end{equation*}

\section{Fitted parameters}

\begin{figure}[!ht]
\includegraphics[width=0.89\columnwidth]{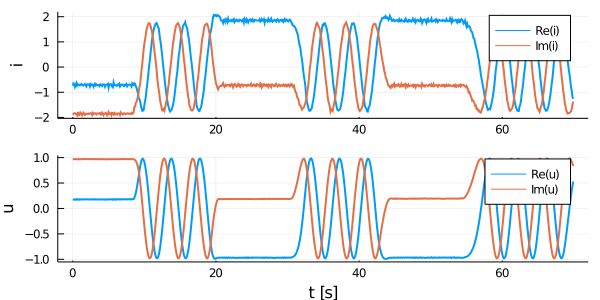}

\caption[]{TECNALIA inverter measurement: These are the measured time series for current and voltage at the grid forming inverter described in \cite{planas2013design}. The oscillations correspond to shifts in the frequency of the power source in the lab setup.}
\label{fig:measurement}
\end{figure}

We get the following parameters by fitting the model as described in section \ref{sec:tecnalia}:

\begin{center}
\begin{tabular}{c|c|c|c|c|c}
    & $A$ & $B$ & $C$ & $G$ & $H$ \\
    \hline
    $\Re(u)$ & 3.9426 & -0.0064 & -4.3700 & -0.1561 & -0.0022 \\
    \hline
    $\Im(u)$ & -0.4919 & 0.8614 & -0.0672 & -0.3988 &  0.6019 \\
    \hline
    $\omega$ & -1.9899 & -0.9445  & -2.0393 & -2.8718 & 4.5051 \\
\end{tabular}
\end{center}

All parameters have the units $1/s$ in the per unit system with $P_{base} = 10 kW$ and $V_{base} = 393.4 V$.

\section{Linear stability (infinite busbar)}\label{sec:linear_stability}

A crucial aspect of obtaining a normal form is that it allows us to make highly general analytic statements that apply directly (if approximately) to a wide range of potential power grid components. To demonstrate this point with a proof of concept, and to further improve our understanding of the coefficients in eqns.~\eqref{eq:normal_form_uj}, we will consider the linear stability when connected to an infinite bus (or slack node) via eqns.~\eqref{eq:ss_algebraic_pq}. We will limit ourselves to the normal form with a single internal frequency variable, i.e. eqns.~\eqref{eq:third_order_app}.

We work in the reference frame co-rotating with the infinite bus and fix its phase angle at zero such that the nodal voltage at the infinite bus is given by a constant $V_s \in \mathbb{R}_{>0}$. Our goal is to derive conditions for the parameters that ensure local asymptotic stability for some valid equilibrium point with synchronized frequency, i.e. some $y_0 = (0, \rho_0^2, p_0, q_0)^T$ for which $A^u(y_0) = A^\omega(y_0) = 0$, and such that $\exists \varphi \in [0,2\pi): p_0 + iq_0 = Y^* (\rho_0^2 - \rho_0 V_s e^{i\varphi})$. Note that, with slight abuse of notation, $Y$ denotes the admittance of the single transmission line here. For convenience we make the change of coordinates $\sigma_R + i\sigma_I := \ln u$, with the subscripts $R,I$ denoting real and imaginary part in the following, and set $B^u_I = 1$ without loss of generality. The system we are considering here is thus given by
\begin{equation}\label{eq:ideal}\begin{aligned}
p + i q &= Y^*(e^{2\sigma_R} - e^{\sigma_R + i\sigma_I} V_s) \;,\\
\dot{\sigma}_R + i\dot{\sigma}_I &= B^u \delta\omega + C^u \delta\rho^2 + G^u \delta p + H^u \delta q \;,\\
\dot{\delta\omega} &= B^{\omega} \delta\omega + C^{\omega} \delta\rho^2 + G^{\omega} \delta p + H^{\omega} \delta q \;,
\end{aligned}\end{equation}
with the Jacobian
\begin{equation} 
J(y_0) = \begin{bmatrix} 2 C^u\rho_0^2 + G^u p_0^+ + H^u q_0^- & H^u p_0^- - G^u q_0^+ & B^u_R \\ 0 & 0 & 1 \\
2 C^\omega \rho_0^2 + G^\omega p_0^+ + H^\omega q_0^- & H^\omega p_0^- - G^\omega q_0^+ & B^\omega \end{bmatrix} \;,
\end{equation}
and the constants $p_0^\pm$, $q_0^\pm$ defined as 
\begin{align*}
	p_0^\pm &:= p_0 \pm Y_R\rho_0^2 \;,\\
	q_0^\pm &:= q_0 \pm Y_I\rho_0^2 \;.
\end{align*}
Invoking the Routh-Hurwitz criterion~\cite{hahn1967} we can ensure all three eigenvalues to lie in the left complex half-plane if the inequalities 
\begin{equation}\label{eq:routh-hurwitz}\begin{aligned}
	\mathrm{tr} J(y_0) &< 0 \;,\\
	\det J(y_0) &< 0 \;,\\
	\mathrm{tr} J(y_0) \left( \mathrm{tr} J(y_0)^2 - \mathrm{tr}^2 J(y_0) \right) &< 2\det J(y_0) \;, 
\end{aligned}\end{equation}
are satisfied (see appendix~\eqref{eq:stability_full} for these inequalities in terms of the parameters). While these conditions are necessary and sufficient, they are too intricate to yield any qualitative insights, so we consider the special case in which the response of the node to active and reactive power is adapted to the behavior of the power line:
\begin{equation}\label{eq:ideal_droop}\begin{aligned}
G^u &= -k^u \cos\kappa \;, \quad H^u = -k^u \sin\kappa \;,\\
G^\omega &= -k^\omega \sin\kappa \;, \quad H^\omega = k^\omega \cos\kappa \;,
\end{aligned}\end{equation}
for some $k^u > 0$, $k^\omega > 0$ and $\tan \kappa := -Y_I/Y_R$. This makes the coupling behave like conventional droop control~\cite{schiffer2014} for a purely inductive network ($Y_R = 0$) even when $Y_R \geq 0$ (an idea which has been used e.g. for dispatchable virtual oscillator control~\cite{doerfler2018}). To state the sufficient stability conditions, we first define the short-hands
\begin{align*}
    R^u := 1 - \frac{C^u}{k^u \lvert Y \rvert} \;,\quad R^\omega := \frac{C^\omega}{k^\omega \lvert Y \rvert} \;,\quad R^V := \frac{V_s}{2\rho_0} \;,
\end{align*}
encoding the ratios between the coefficients specifying the system's reaction on voltage amplitude and power deviations, as well as the ratio between the infinite bus voltage and the desired voltage amplitude at the node (although in practical cases we usually have $R^V \approx 1/2$). Additionally, we define the angle
\begin{equation*}
    \gamma := \tan^{-1}\left( \frac{R^\omega}{R^u}\right) \;.
\end{equation*}
With these definitions and the assumption~\eqref{eq:ideal_droop}, we can state that the system~\eqref{eq:ideal} is asymptotically stable if the following conditions are satisfied:
\begin{equation*}
    \lvert \varphi \rvert \leq \frac{\pi}{2} \;,\quad B^\omega < 0 \;,\quad \mathrm{sign}(\varphi)B^u_R \geq 0 \;,
\end{equation*}
\begin{equation}\label{eq:stability_condition1}\begin{aligned}
    R^u &> R^V \cos\varphi \;,\\
    \mathrm{sign}(\varphi) R^\omega &\leq R^V \lvert \sin\varphi \rvert \;,
\end{aligned}\end{equation}
\begin{equation}\label{eq:stability_condition2}
    \cos (\varphi - \gamma) \sqrt{(R^u)^2 + (R^\omega)^2} > R^V \;.
\end{equation}

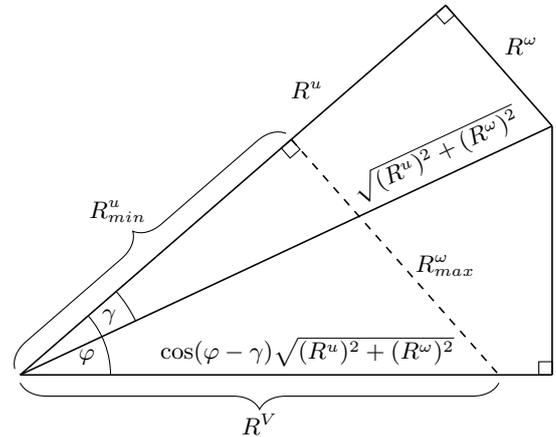
\begin{figure}
\begin{tikzpicture}[scale=0.7]
    \coordinate (O) at (0,0);
    \coordinate (BR) at (10,0);
    \coordinate (TR) at (10,33/7);
    \coordinate (T) at (8,7);
    \coordinate (B) at (9,0);
    \coordinate (TL) at (9*64/113,9*56/113);
    
    \tkzDrawLines[add=0 and 0](O,BR) 
    \tkzDrawLines[add=0 and 0](TR,T)
    \tkzDrawLines[add=0 and 0](T,O) 
    \tkzDrawLines[add=0 and 0](TR,O)
    \tkzDrawLines[add=0 and 0](BR,TR)
    \tkzDrawLines[dashed, add=0 and 0](TL,B)
    
    \tkzLabelLine[pos=0.75, left, shift={(-0.1,0.1)}](O,T){$R^u$}
    \tkzLabelLine[pos=0.4, right, shift={(0.1,0.1)}](T,TR){$R^\omega$}
    \tkzLabelLine[pos=0.8, above, sloped](O,TR){$\sqrt{(R^u)^2 + (R^\omega)^2}$}
    \tkzLabelLine[pos=0.65, above, shift={(0.25,0.1)}](TL,B){$R^\omega_{max}$}
    \tkzLabelLine[pos=0.54, above, sloped](O,BR){$\cos(\varphi-\gamma)\sqrt{(R^u)^2 + (R^\omega)^2}$}
    \draw [decorate,decoration={brace,amplitude=10pt, mirror, raise=3}] (O) -- node [below, shift={(0,-0.4)}] {$R^V$} (B);
    \draw [decorate,decoration={brace,amplitude=10pt, raise=3}]
    (O) -- node [above, shift={(-0.5,0.35)}] {$R^u_{min}$} (TL);
    
    \tkzMarkRightAngle(O,BR,TR);
    \tkzMarkRightAngle(O,T,TR);
    \tkzMarkRightAngle(O,TL,B);
    
    \tkzMarkAngle[arc=l,size=1.7,mark=](B,O,T);
    \tkzMarkAngle[arc=l,size=2.4,mark=](TR,O,T);
    \node at ($(O)+(15:13mm)$) {$\varphi$};
    \tkzLabelAngle[pos=2](TR,O,T){$\gamma$};
\end{tikzpicture}
\caption{Geometric representation of inequalities~\eqref{eq:stability_condition1} and~\eqref{eq:stability_condition2} for $\mathrm{sign}(\varphi) = \mathrm{sign}(R^\omega)$, with $R^u_{min}:= R^V\cos\varphi$ and $R^\omega_{max}:= R^V\lvert\sin\varphi\rvert$}
\label{fig:stability_conditions}
\end{figure}

The derivation of these inequalities can be found in appendix~\ref{sec:app_linear_stability}. The conditions show that the chief determinants for the stability of a certain equilibrium (with relative phase angle $\varphi$ and voltage amplitude $\rho_0$) are the ratios between the coefficients specifying the system's reaction on voltage amplitude and power deviations. A geometric representation of the inequalities~\eqref{eq:stability_condition1} and~\eqref{eq:stability_condition2} is depicted in fig.~\ref{fig:stability_conditions}. Assuming $V_s \approx \rho_0$ (as is usually the case for power grids), it can be seen that for relative angles $\lvert\varphi\rvert < \pi/3$, it is enough to ensure the correct sign for the coefficients $C^u$ and $C^\omega$, i.e. $C^u < 0$ and $\mathrm{sign}(\varphi) C^\omega \geq 0$, while respecting the bound on $R^\omega$. $R^\omega$ parametrizes the impact of the amplitude on the frequency of the system, relative to the power droop. Thus we find that in a moderately loaded scenario, the crucial factor for the linear stability of the system is the amplitude-phase coupling.


If a greater load, and thus a greater relative phase angle needs to be guaranteed stable, we must further ensure that the influence of voltage amplitude deviations dominates that of power deviations, as given by inequality~\eqref{eq:stability_condition2}. Lastly, we note that the inequality~\eqref{eq:stability_condition2} is actually a necessary bound under the assumption of eqns.~\eqref{eq:ideal_droop}. The inequalities~\eqref{eq:stability_condition1} are only sufficient, i.e. they may be relaxed by invoking stricter bounds on $B^\omega$ and $B^u_R$, which are rather technical however (see eqn.~\ref{eq:stability_simpler3}). 

Now, given a concrete model from the class of amplitude-frequency oscillators, we only have to substitute the coefficients with the corresponding partial derivatives as given by eqns.~\eqref{eq:normal_form} in order to translate the stability conditions to the specific model parameters. 

\section{Numerical simulations} \label{sec:numerics_appendix}

This appendix describes the model and simulation setup for section~\ref{sec:validation}, as well as some further results for the networked case. All simulations were performed using DifferentialEquations.jl \cite{rackauckas2017differentialequations}. The network model was built with PowerDynamics.jl \cite{plietzsch2021powerdynamics}. The simulations were performed with a RADAU solver \cite{wanner1996solving} with relative tolerances set to $10^-6$ for more details we refer to the code accompanying the paper.

\begin{figure}[!ht]
\includegraphics[width=0.69\columnwidth]{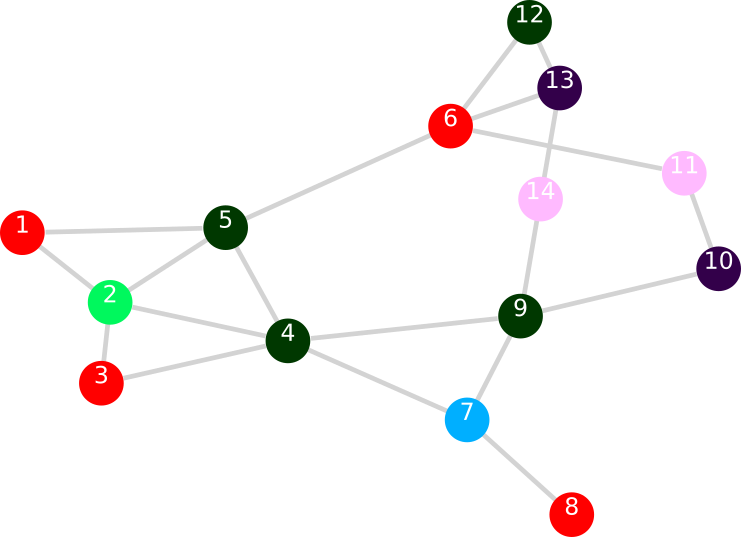}

\caption[]{IEEE 14-bus test system~\cite{canizares2003modeling}: light green is the slack bus, red are synchronous machines~\eqref{eq:schmietendorf}, dark green are inverters~\eqref{eq:schiffer}, blue is a passive node, purple are inverters~\eqref{eq:dVOC}, pink are inverters~\eqref{eq:schiffer_inertia}.}
\label{fig:IEEE14_net}
\end{figure}

\subsection{Infinite busbar}

In the inifinite bus simulations of section~\ref{sec:numerics-inf-bus} we consider a model of a droop-controlled inverter, eqns.~\eqref{eq:schiffer}, and a third-order approximation of a synchronous machine, eqns.~\eqref{eq:schmietendorf}, connected to an infinite bus. We use pu units with voltages $1$ at the nodes and a per unit power chosen such that the admittance of the line, which we take to be purely inductive, is $Y = -1i$. The line model also includes shunt capacitance of $Y^s = 0.2i$. For the inverter model~\eqref{eq:schiffer} we choose the time constant $\tau_p = 2.5$ and droop gains $k_p = 5$, $k_q = 0.1$, and expand around $\rho_0 = V^d = 1$, $\omega_0 = \omega^d = 0$, $p_0 = p^d = 0.5$ and $q_0 = q^d = 0.2$. For the synchronous machine model~\eqref{eq:schmietendorf} we choose the damping constant $\gamma=0.2$, time constant $\alpha=2$ and internal reactance $X=1$, and expand around $\rho_0 = E^f = 1$, $\omega_0 = 0$ (the model is given in the co-rotating reference frame), $p_0 = p^m = 0.5$ and $q_0 = 0$ as no explicit set-point for reactive power is given in the model.

\begin{figure*}[!ht]
\includegraphics[width=0.3\textwidth]{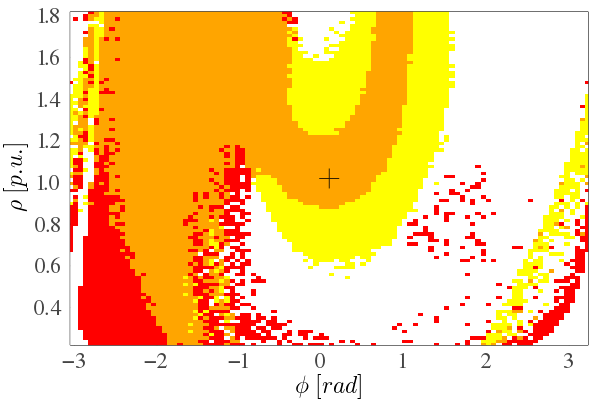}
\includegraphics[width=0.3\textwidth]{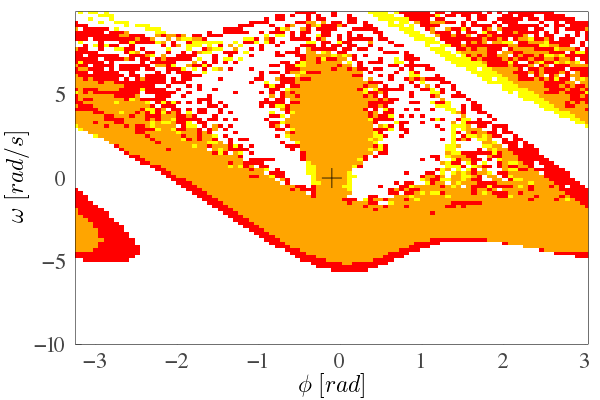}
\includegraphics[width=0.3\textwidth]{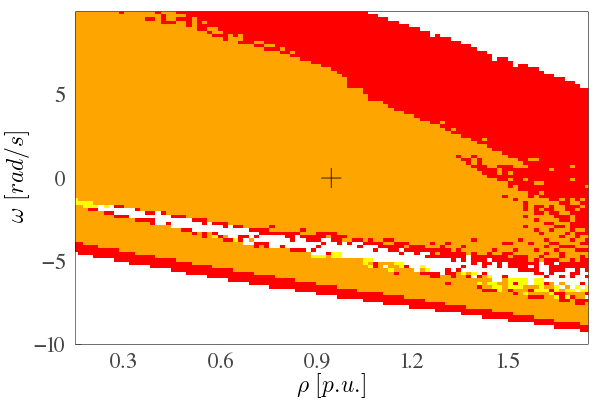}

\includegraphics[width=0.3\textwidth]{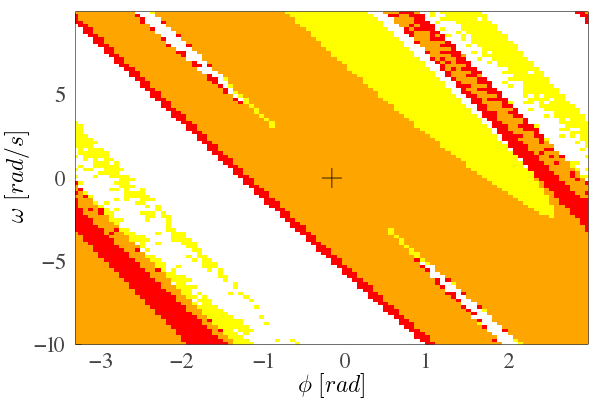}
\includegraphics[width=0.3\textwidth]{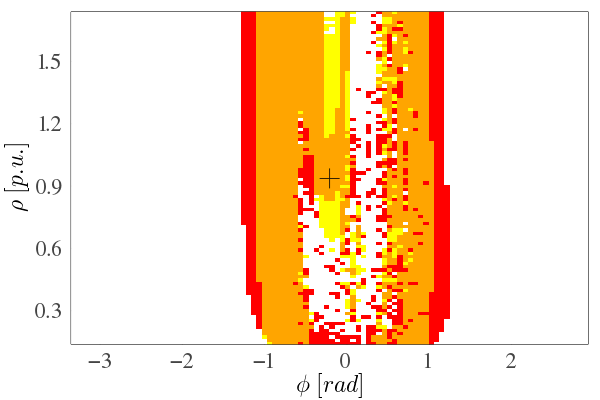}
\includegraphics[width=0.3\textwidth]{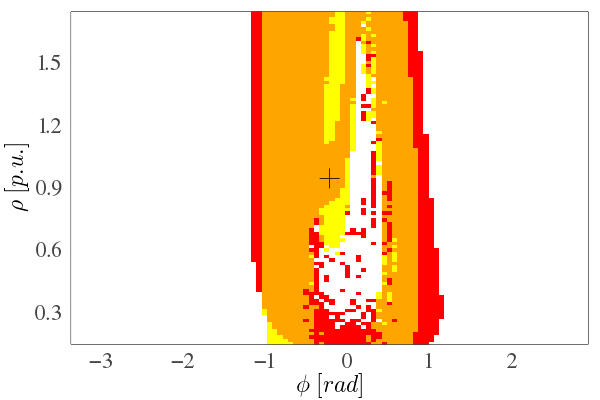}
\caption[]{Various basin slices for the IEEE 14-bus test case. Color code as in Fig.~\ref{fig:infinite_busbar_results}.
First row: bus 1 (left), bus 4 (middle), bus 6 (right). Second row: bus 8 (left), bus 9 (middle), bus 12 (right)}
\label{fig:IEEE14_slices}
\end{figure*}

\subsection{Network}

The network model is based on the IEEE-14 bus dynamical test system, augmented with grid forming components. The distribution of components is shown in Figure~\ref{fig:IEEE14_net}. We use are synchronous machines at nodes 1, 3, 6 and 8 (eqns.~\eqref{eq:schmietendorf}) and inverters with different types of control at most other nodes, namely at 4, 5, 9, and 12 we use the droop controlled inverter~\eqref{eq:schiffer}, at nodes 11 and 14 we attach the inverter model~\eqref{eq:schiffer_inertia} (which adds inertia to the amplitude dynamics), and at buses 10 and 13 we place inverters with dispatchable virtual oscillator control~\eqref{eq:dVOC}. For the detailed parameter choices we refer the reader to the code.

To stress the system we scale the active and reactive power demand at all inverter controlled nodes by a common factor $f_s$ varying between $0$ and $2$. Note that the set points do not provide a solution of the power flow, and we have persistent deviations from the desired operating points.

The main part of the paper discusses the single node basin stability for a variety of $f_s$ values. Here we also briefly present more phase space slices in the style of Figure~\ref{fig:infinite_busbar_results}. Each location in the slice corresponds to an initial condition for two variables of the system. All other variables in the system are intialized at the quasi-steady operating state of the network. Thus the plots show very large, instantaneous perturbations affecting only a single node. The results are shown in fig.~\ref{fig:IEEE14_slices}.

We see that in this highly challenging scenario with large perturbations, the agreement with respect to the shape of the stability region becomes significantly worse the further we stray from the operating state. Most problematic appear large deviations in voltage amplitude. This should be kept in mind, when analyzing fault scenarios that feature such deviations. However, many important qualitative features are still captured by the normal form.

\section{Coordinate transformation}\label{sec:appendix_transformation}

Here we write out the steps that lead from eqns.~\eqref{eq:1-oscillator} to eqns.~\eqref{eq:transformed}. First we define
\begin{align*}
    &\tilde{f}^{u,z,x}(u, \xi, \rho^2, p, q) := \\
    &f^{u,z,x} \left( x, \frac{u (\psi - i\chi)}{\rho^2}, \frac{\psi + i\chi}{u}, u, \frac{\rho^2}{u}, \frac{u (p - iq)}{\rho^2}, \frac{p + iq}{u} \right) \;.
\end{align*}
Then we have
\begin{align*}
    \dot{\psi} + i\dot{\chi} =& ~\dot{u} z^* + u\dot{z}^* = \\
    &\tilde{f}^u \frac{\psi + i\chi}{u} + u (\tilde{f}^z)^* =: \tilde{f}^\psi + i\tilde{f}^\chi \;,
\end{align*}
and define
\begin{equation*}
    \tilde{f}^\xi := \begin{pmatrix} \tilde{f}^x \\ \tilde{f}^\psi \\ \tilde{f}^\chi \end{pmatrix} \;.
\end{equation*}

\section{Linear time invariant input-output form of the internal dyanmics}\label{sec:appendix_mimo}

We can write \eqref{eq:normal_form_uj} in terms of real variables $x$, $x^i$ and $x^o$, and real matrices $A^M$, $B^M$, $C^M$, $D^M$:

\begin{equation}\label{eq:normal_form_mimo}\begin{aligned} 
	&\delta p + i\delta q = u j^* - (p_0 + i q_0)\\
	&\delta\rho^2 = u u^* - \rho_{0}^2 \;,\\
	&\frac{\dot{u}}{u} = x^o_1 + i x^o_2 \;,\\
    &x^i =  [\delta\rho^2, \delta p, \delta q, 1]\\
	&\dot{x} = A^{M} x + B^{M} x^i\\
    &x^o = C^{M} x + D^{M} x^i
\end{aligned}\end{equation}

This form is the most suitable for using tools from the study of LTI systems for power grid models.

\section{Local asymptotic stability}\label{sec:app_linear_stability}

The full inequalities~\eqref{eq:routh-hurwitz} in terms of the normal form coefficients and the expansion point are given by the rather lengthy expressions
\begin{equation}\label{eq:stability_full}\begin{aligned}
0 >& ~2\rho_0^2 C^u + G^u p_0^+ + H^u q_0^- + B^\omega \;,\\[5pt]
0 <& ~2\rho_0^2 \left( (H^\omega p_0^- - G^\omega q_0^+) C^u - (H^u p_0^- - G^u q_0^+) C^\omega \right)& \\
&- (p_0^+ p_0^- + q_0^+ q_0^-) (H^u G^\omega - G^u H^\omega) \;,\\[5pt]
0 <& -(2\rho_0^2 C^u + G^u p_0^+ + H^u q_0^-)^2 B^\omega \\
&- (2\rho_0^2 C^u + G^u p_0^+ + H^u q_0^-) (B^\omega)^2 \\ &+(H^\omega p_0^- - G^\omega q_0^+) B^\omega \\ 
&+ B^u_R (2\rho_0^2 C^u + G^u p_0^+ + H^u q_0^- + B^\omega) \times \\
& \times (2\rho_0^2 C^\omega + G^\omega p_0^+ + H^\omega q_0^-) \\
&+ (2 C^\omega \rho_0^2 + G^\omega p_0^+ + H^\omega q_0^-) ( H^u p_0^- - G^u q_0^+) \;. 
\end{aligned}\end{equation}
Employing the additional assumptions of eqns.~\eqref{eq:ideal_droop} and the definitions of $R^u$, $R^\omega$, and $R^V$, yields the more compact inequalities
\begin{equation}\label{eq:stability_simpler1}
    0 > ~W_1 + B^\omega \;,
\end{equation}
\begin{equation}\label{eq:stability_simpler2}
    0 > - R^u \cos\varphi - R^\omega \sin\varphi + R^V \;,
\end{equation}
\begin{equation}\label{eq:stability_simpler3}\begin{aligned}
    0 >& ~2\rho_0^2 \lvert Y \rvert (k^u)^2 B^\omega W_1^2 + k^u (B^\omega)^2 W_1 \\
    &- k^\omega B^u_R W_2 \left( 2\rho_0^2 \lvert Y \rvert k^u W_1 + B^\omega \right) \\
    &+ k^\omega B^\omega R^V \cos\varphi + 2\rho_0^2 \lvert Y \rvert k^u k^\omega R^V W_2 \sin\varphi
\end{aligned}\end{equation}
with
\begin{align*}
    W_1 &:= R^V \cos\varphi - R^u \;,\\
    W_2 &:= R^\omega - R^V \sin\varphi \;.
\end{align*}


From these inequalities we can immediately deduce the conditions given in section~\ref{sec:linear_stability}. By requiring $B^\omega <0$, inequality~\eqref{eq:stability_simpler1} is satisfied if $W_1 < 0$, which yields the first of inequalities~\eqref{eq:stability_condition1}. By further requiring $\lvert \varphi \rvert \leq \pi/2$, inequality~\eqref{eq:stability_simpler3} is satisfied if
\begin{align*}
    0 &\geq W_2 \sin\varphi \;,\\
    0 &\leq B^u_R W_2 \;,
\end{align*}
which is equivalent to $\mathrm{sign}(\varphi) B^u_R \geq 0$ and the second of inequalities~\eqref{eq:stability_condition1}. For inequality~\eqref{eq:stability_simpler2} we make use of the trigonometric identity
\begin{equation*}
    a \cos\alpha + b \sin\alpha = c \cos(\alpha + \beta) \;,
\end{equation*}
with
\begin{equation*}
    c := \mathrm{sign}(a) \sqrt{a^2 + b^2} \;,\quad \beta := \tan^{-1}\left( -\frac{b}{a} \right) \;.
\end{equation*}
Since $W_1 < 0$ implies $R^u >0$, this yields inequality~\eqref{eq:stability_condition2}.
\\

\section{Models used in section~\ref{sec:validation}}\label{sec:appendix_models}

For completeness we give the models we base the heterogeneous network in section~\ref{sec:validation} on. Here the parameters are kept in line with the notation used in the original papers that they were taken from. We also provide the normal form coefficients when expanded around the design set-points.\\

Third-order approximation of synchronous machines~\cite{schmietendorf2013}:
\begin{equation}\label{eq:schmietendorf}\begin{aligned}
    \ddot{\phi}_n &= -\gamma_n \dot{\phi}_n + p_n^m - p_n \\
    \alpha_n\dot{E}_n &= E^f_n - E_n - X_n \frac{q_n}{E_n}
\end{aligned}\end{equation}

    \begin{center}
    \begin{tabular}{c|c|c|c|c|c}
        & $A$ & $B$ & $C$ & $G$ & $H$ \\
        \hline
        $u$ & $0$ & $i$ & $-\frac{1}{2\alpha_n (E^f_n)^2}$ & $0$ & $-\frac{X_n}{\alpha_n (E^f_n)^2}$    \\
        \hline
        $\omega$ & $0$ & $-\gamma_n$ & $0$ & $-1$ & $0$ \\
    \end{tabular}
    \end{center}

\phantom{A}\\
Droop-controlled inverter~\cite{schiffer2014}:
\begin{equation}\label{eq:schiffer}\begin{aligned}
    \dot{\phi}_n &= \omega_n \\
    \tau_{p_n}\dot{\omega}_n &= -\omega_n + \omega^d - k_{p_n} (p_n - p_n^d) \\
    \tau_{p_n}\dot{V}_n &= -V_n + V_n^d - k_{q_n} (q_n - q_n^d)
\end{aligned}\end{equation}

    \begin{center}
    \begin{tabular}{c|c|c|c|c|c}
        & $A$ & $B$ & $C$ & $G$ & $H$ \\
        \hline
        $u$ & $i\omega^d$ & $i$ & $-\frac{1}{2\tau_{p_n} (V^d_n)^2}$ & $0$ & $-\frac{k_{q_n}}{\tau_{p_n} V^d_n}$     \\
        \hline
        $\omega$ & $0$ & $-\frac{1}{\tau_{p_n}}$ & $0$ & $-\frac{k_{p_n}}{\tau_{p_n}}$ & $0$ \\
    \end{tabular}
    \end{center}

\phantom{A}\\
Droop-controlled inverter~\cite{schiffer2014} without the assumption of near instantaneous voltage measurement:
\begin{equation}\label{eq:schiffer_inertia}\begin{aligned} 
	\dot{\phi}_n &= \omega_n \\
    \tau_{P_n}\dot{\omega}_n &= -\omega_n + \omega^d - k_{P_n} (P_n - P_n^d) \\
    \tau_{P_n}\tau_{V_n}\ddot{V}_n &= -(\tau_{P_n} + \tau_{V_n}) \dot{V}_n - V_n + V_n^d - k_{Q_n} (Q_n - Q_n^d)
\end{aligned}\end{equation}

    Writing $\omega$ for $\xi_1$, $\nu$ for $\xi_2$, $B^\times_\omega$ for $B^\times_1$, and $B^\times_\nu$ for $B^\times_2$ we have
    \begin{center}
    \begin{tabular}{c|c|c|c|c|c|c}
        & $A$ & $B_\omega$ & $B_\nu$ & $C$ & $G$ & $H$ \\
        \hline
        $u$ & $i\omega^d$ & $i$ & $1$ & $0$ & $0$ & $0$     \\
        \hline
        $\omega$ & $0$ & $-\frac{1}{\tau_{p_n}}$ & $0$ & $0$ & $-\frac{k_{p_n}}{\tau_{p_n}}$ & $0$  \\
        \hline
        $\nu$ & $0$ & $0$ & $\frac{\tau_{P_n}+\tau_{V_n}}{\tau_{P_n}\tau_{V_n}}$ & $-\frac{1}{2\tau_{P_n}\tau_{V_n} (V^d_n)^2}$ & $0$ & $-\frac{k_{q_n}}{\tau_{P_n}\tau_{V_n} V^d_n}$
    \end{tabular}
    \end{center}

\phantom{A}\\
Dispatchable virtual oscillator control~\cite{doerfler2019}:
\begin{equation}\label{eq:dVOC}\begin{aligned}
    \dot{u}_n = &\left( \alpha\eta + i\omega_0 + \frac{\eta e^{i\kappa}}{(v^\star_n)^2}(p^\star_n -iq^\star_n) \right) u_n \\
    &-\frac{\alpha\eta}{(v^\star_n)^2} u_n \lvert u_n \rvert^2 -\eta e^{i\kappa} j_n
\end{aligned}\end{equation}

    \begin{center}
    \begin{tabular}{c|c|c|c|c}
        & $A$ & $C$ & $G$ & $H$ \\
        \hline
        $u$ & $i\omega_0$ & $-\frac{\alpha\eta}{(v^\star_n)^2} + \frac{\eta e^{i\kappa}}{(v^\star_n)^4}(p^\star_n -iq^\star_n)$ & $-\frac{\eta e^{i\kappa}}{(v^\star_n)^2}$ & $\frac{i\eta e^{i\kappa}}{(v^\star_n)^2}$     \\
        \end{tabular}
    \end{center}

\end{document}